\documentclass[10pt,twocolumn,twoside]{IEEEtranTCOM}

\ifCLASSINFOpdf
\else
\fi
\usepackage[linesnumbered,ruled,vlined]{algorithm2e}
\usepackage{enumerate}
\usepackage{graphicx}
\usepackage{enumitem}
\usepackage{amsmath}
\usepackage{algpseudocode}
\usepackage{amssymb}
\usepackage{times}
\usepackage{endnotes}
\usepackage{stfloats}
\usepackage{stkernel}
\usepackage{color}
\usepackage{enumerate}
\usepackage{array}
\usepackage{relsize}
\usepackage[belowskip=-10pt,aboveskip=3pt,footnotesize]{caption}
\usepackage{cite}
\renewcommand{\baselinestretch}{0.98}

\newcommand{\N}{\ensuremath{{\mathcal{N}}}}
\newcommand{\C}{\ensuremath{{ \mathcal{C} }}}

\newcommand{\bm}{\mathbf}

\def\argmax{\mathop{\rm arg\,max}}
\newcommand{\refeq}[1]{(\ref{#1})}
\newcommand{\tr}[1]{\textrm{#1}}
\newcommand{\Pdtheta} {p_{\tr d}^{(\theta)} }

\newcommand{\Pftheta} {p_{\tr f}^{(\theta)} }
\newcommand{\Pbtheta} {p_{\tr b}^{(\theta)} }
\newcommand{\Puc} {P_{\tr u}^{(c)} }
\newcommand{\Pdc} {P_{\tr d}^{(c)} }
\newcommand{\Pdelta} {p_{\Delta} }

\newcommand{\bbTheta}{\boldsymbol{\bar{\Theta}}}
\newcommand{\bThetak}[1]{\boldsymbol{\Theta}_{#1}}

\newcommand{\thetak}[2]{\theta_{{ #1}}^{{ (#2)}}}

\newcommand{\Deltakn}[2]{\Delta_{{ #1}}^{{ (#2)}}}

\newcommand{\bThetahatk}[1]{\widehat{\boldsymbol{\Theta}}_{#1}}

\newcommand{\thetahatk}[2]{\hat{\theta}_{{ #1}}^{{ (#2)}}}

\newcommand{\bbr}{\bm{\bar{r}}}
\newcommand{\brk}[1]{\bm{r}_{#1}}
\newcommand{\rk}[2]{{r}_{{ #1}}^{{ (#2)}}}

\newcommand{\sigmap}[2]{\sigma_{{ #1}}^{{ #2} }}

\newcommand{\bbc}{\bm{\bar{c}}}
\newcommand{\bck}[1]{\bm{c}_{#1}}
\newcommand{\ck}[2]{{c}_{{ #1}}^{{ (#2)}}}

\newcommand{\bwk}[1]{\bm{w}_{#1}}
\newcommand{\wk}[2]{{w}_{{ #1}}^{{ (#2)}}}

\newcommand{\alphak}[2]{\alpha_{{ #1}}^{{ (#2)}}}
\newcommand{\betak}[2]{\beta_{{ #1}}^{{ (#2)}}}
\newcommand{\afk}[2]{a_{\textrm{f}, #1}^{ (#2)}}
\newcommand{\bafk}[2]{\bar{a}_{\textrm{f}, #1}^{ (#2)}}
\newcommand{\bbafk}[2]{\bar{\bar{a}}_{\textrm{f}, #1}^{ (#2)}}
\newcommand{\tafk}[2]{\tilde{a}_{\textrm{f}, #1}^{ (#2)}}
\newcommand{\ttafk}[2]{\tilde{\tilde{a}}_{\textrm{f}, #1}^{ (#2)}}
\newcommand{\tabk}[2]{\tilde{a}_{\textrm{b}, #1}^{ (#2)}}
\newcommand{\ttabk}[2]{\tilde{\tilde{a}}_{\textrm{b}, #1}^{ (#2)}}
\newcommand{\abk}[2]{a_{\textrm{b}, #1}^{ (#2)}}
\newcommand{\babk}[2]{\bar{a}_{\textrm{b}, #1}^{ (#2)}}
\newcommand{\bbabk}[2]{\bar{\bar{a}}_{\textrm{b}, #1}^{ (#2)}}
\newcommand{\xk}[2]{x_{ #1}^{ (#2)}}
\newcommand{\yk}[2]{y_{ #1}^{ (#2)}}
\newcommand{\zk}[2]{z_{ #1}^{ (#2)}}
\newcommand{\tzk}[2]{\tilde{z}_{ #1}^{ (#2)}}
\newcommand{\uk}[2]{u_{ #1}^{ (#2)}}
\newcommand{\utk}[2]{\tilde{u}_{ #1}^{ (#2)}}

\newcommand{\uttk}[2]{\tilde{\tilde{u}}_{ #1}^{ (#2)}}
\newcommand{\ytk}[2]{\tilde{y}_{ #1}^{ (#2)}}

\newtheorem{remark}{Remark}

\begin{document}

\title{Algorithms for Joint Phase Estimation and Decoding for MIMO Systems in the Presence of Phase Noise}

\author{Rajet~Krishnan,~\IEEEmembership{Student~Member,~IEEE,}~Giulio~Colavolpe,~\IEEEmembership{Senior~Member,~IEEE} ~Alexandre~Graell~i~Amat,~\IEEEmembership{Senior~Member,~IEEE,}~and~Thomas~Eriksson

\thanks{Rajet Krishnan, Alexandre Graell i Amat, and Thomas Eriksson are with the Department
of Signals and Systems, Chalmers University of Technology, Gothenburg, Sweden (e-mail: \{rajet,
alexandre.graell, thomase\}@chalmers.se).}%
\thanks{Giulio Colavolpe is with Dipartimento di Ingegneria dell'Informazione,  University of Parma, Parma, Italy  (e-mail: giulio@unipr.it).}
\thanks{Research supported by the Swedish Research Council under grant \#2011-5961.}

}%

\markboth{IEEE Transactions on Signal Processing}%
{Submitted paper}

\maketitle

\begin{abstract}
In this work, we derive the maximum a posteriori (MAP) symbol detector for a multiple-input multiple-output system in the presence of Wiener phase noise due to noisy local oscillators. As in single-antenna systems, the computation of the optimal receiver is an infinite dimensional problem and is thus unimplementable in practice. In this purview, we propose three suboptimal, low-complexity algorithms for approximately implementing the MAP symbol detector, which involve joint phase noise estimation and data detection. Our first algorithm is obtained by means of the sum-product algorithm, where we use the multivariate Tikhonov canonical distribution approach. In our next algorithm, we derive an approximate MAP symbol detector based on the smoother-detector framework, wherein the detector is properly designed by incorporating the phase noise statistics from the smoother. The third algorithm is derived based on the variational Bayesian framework. By simulations, we evaluate the performance of the proposed algorithms for both uncoded and coded data transmissions, and we observe that the proposed techniques significantly outperform the other algorithms proposed in the literature.


\end{abstract}

\section{Introduction}
\label{sec:intro}

\PARstart{E}{mploying} multiple-input multiple-output (MIMO) systems  has been shown to significantly enhance performance in terms of data rate and link reliability in wireless fading environments \cite{paulraj}. In general, the analysis and design of MIMO system is based on the assumption that the carrier phase is perfectly known at the receiver, and that there is no phase noise in the system. The phase noise manifests in a MIMO system as the random, time-varying phase differences between the oscillators connected to the antennas at the transmitter and the receiver. Practical designs of MIMO systems based on this assumption can result in significant performance losses and have to be addressed appropriately \cite{baum11}. The detrimental effects of phase noise can be even more pronounced in scenarios where independent oscillators are connected to each transmit and receive antenna (or a subset of them). This scenario is particularly relevant for line-of-sight MIMO systems that operate at carrier frequencies of around $10$ GHz or lesser. Here, separate oscillators are needed for each antenna \cite{meyr05}, since the antennas are placed far from each other \cite{Øien07}. The scenario under consideration also corresponds to a massive MIMO system \cite{marzetta10, larsson12}, where a large number of antennas are placed at the base station and each user terminal is equipped with a single antenna. 

The problem of designing receiver algorithms in the presence of random, time-varying phase noise due to noisy local oscillators has been studied extensively for single-antenna systems. We refer the readers to \cite{marc99,rajet13,rajet13_1,colavolpe05,nissila09,noels03} and the references therein. To address the problem of designing receiver algorithms for joint phase noise estimation and data detection, the expectation-maximization (EM) framework is applied in \cite{noels03}, resulting in a code-aided synchronization technique. In \cite{colavolpe05}, receiver algorithms are developed based on the sum-product algorithm (SPA) by constraining the probability density functions (pdfs) computed by the SPA to be in a certain canonical family (for e.g., the exponential family). This method of constraining the pdfs is referred to as the canonical distribution approach \cite{worthen01}, and in particular, using the Tikhonov canonical distribution in \cite{colavolpe05} results to be the most convenient and effective choice. The variational Bayesian (VB) framework is adopted in \cite{nissila09} to develop efficient algorithms for joint phase noise estimation and data detection. In \cite{rajet13}, receiver algorithms are derived by using a smoother-detector structure based on the maximum a posteriori (MAP) symbol detector derived in \cite{kam94}, where the detector is properly designed by incorporating the phase noise statistics from the smoother.


The effect of phase noise on MIMO systems has been investigated in some recent work \cite{hani12, larsson12, baum11, molisch05}, where the impact of phase noise on the MIMO channel measurements and the estimated capacity is studied. In \cite{meyr05}, data-aided estimation of phase noise is studied using a Wiener filter. In \cite{hani12}, the problem of joint channel and phase noise estimation in a MIMO system is explored, and bounds on the estimation performance are derived. Soft-symbol aided estimation using an extended Kalman Smoother (EKS) and relevant estimation bounds are investigated in \cite{ali13}. However, these works do not consider the problem of designing receiver algorithms for joint phase noise estimation and data detection.  One of the few works investigating this problem can be found in \cite{rajet12}, where the VB framework is employed. In general, MIMO receiver design has focused on developing algorithms for joint channel estimation and data detection (refer to \cite{marc06, haykin06} and the references therein)---it is perceived that the phase noise can be handled by existing channel estimation-data detection algorithms since it can be regarded to be a part of the channel \cite{baum11}.

In this paper, we consider the problem of designing receiver algorithms for joint phase noise estimation and data detection in a MIMO system, where each transmit and receive antenna is connected to an independent noisy oscillator. We focus on the scenario where the phase noise process is a discrete Wiener process \cite{demir00,reza13_1} and drifts much faster than the channel process \cite{rajet12, larsson12}. This implies that the phase noise in the system cannot be handled by moving it into the channel matrix and then compensating it by means of channel estimation---this is a typical scenario when noisy oscillators are used in the system. 

For the MIMO system under consideration, we derive the MAP symbol detector which minimizes the symbol error probability. This receiver structure explicitly involves the estimation of the a posteriori pdf of the phase noise and data detection. The computation of the a posteriori phase noise pdf is an infinite dimensional problem, since the pdf is continuous for the Wiener phase noise process under consideration. This motivates the need for investigating practical, low complexity receiver algorithms for joint phase noise estimation and data detection that also have a good performance. To this end, we propose three new algorithms based on the sum-product algorithm (SPA), the smoother-detector framework from \cite{kam94,rajet13}, and the VB framework in \cite{nissila09}, respectively, for arbitrary number of transmit and receive antennas. We evaluate the performance of the proposed algorithms in strong phase noise scenarios in the presence of Rayleigh fading. We consider both uncoded and coded data transmissions, and compare the performance of the proposed algorithms with that of those available in the literature. We observe that the proposed algorithms significantly outperform those available in the literature.

The remainder of the paper is as follows. In Section \ref{sec:sys_mod}, the MIMO system model under study is presented. We derive the optimal MAP symbol detector in Section \ref{sec:sbs_map}. In Sections \ref{sec:spa}, \ref{sec:estdet}, and \ref{sec:vb}, we derive the SPA-based, smoother-detector-based, and VB-based algorithms, respectively. We present our simulation results in Section \ref{sec:sim_res}. Finally, we summarize our key findings in Section \ref{sec:concl}.

Notation: the expectation and variance operators are denoted as $\mathbb{E}[\cdot]$ and $\mathrm{Var}(\cdot)$, respectively. The conjugate of a complex number is denoted as $[\cdot]^{*}$. $\Re\{\cdot\}$, $\Im\{\cdot\}$, $| \cdot |$, and $\angle\cdot$ are the real, imaginary part, magnitude, and angle of a complex number, respectively. The pdf and probability mass function (pmf) of a random variable are denoted as $p(\cdot)$ and $P(\cdot)$, respectively.

\section{System Model}
\label{sec:sys_mod}

Consider a MIMO system with $N_{\tr t}$ transmit antennas and $N_{\tr r}$ receive antennas. Each antenna is equipped with an independent free-running oscillator that is perturbed by a random phase noise process~\cite{colavolpe05}.  The channel between the transmit-receive antennas is assumed to be known (i.e., estimated), and the phase noise process is assumed to be much faster than the channel~\cite{rajet12}. Data is transmitted as frames consisting of $L$ symbols, and we consider both coded and uncoded transmission.

Assuming Nyquist pulses for transmission, matched filtering followed by sampling at symbol period $T_{\tr s}$, the received signal model in the $k$th time instant at the $n$th receive antenna is
\begin{IEEEeqnarray}{rCl}\label{eq:received_mimo}
\rk{k}{n} &=& \sum_{m= 1}^{N_{\tr t}} \ck{k}{m} e^{\jmath (\thetak{\tr{t},k}{m} + \thetak{\tr{r},k}{n})} + \wk{k}{n} \nonumber\\
&\triangleq& \sum_{m= 1}^{N_{\tr t}} \ck{k}{m} e^{\jmath \thetak{k}{m,n}} + \wk{k}{n},
 \end{IEEEeqnarray}
where perfect timing and frequency synchronization is assumed~\cite{colavolpe05}. Note that in \refeq{eq:received_mimo}, we consider unit channel gains for notational convenience, and the extension of the algorithms developed in the ensuing sections to arbitrary, but known, channel gains is straightforward. In \refeq{eq:received_mimo}, $\ck{k}{m} \in \mathcal{C}$ is the symbol transmitted from the $m$th transmit antenna at the $k$th time instant and drawn equiprobably from an $M$-ary signal constellation set $\mathcal{C}$, and $\wk{k}{n} \sim \N(0,N_{0})$ denotes the zero-mean additive white Gaussian noise (AWGN) at the $n$th receive antenna. The phase noise in the $(m,n)$th link, $\thetak{k}{m,n}$, is defined as the sum of the discrete Wiener phase noise process from the oscillators connected to the $m$th transmit and the $n$th receive antenna, respectively, at time instant $k$,  i.e., $\thetak{k}{m,n} \triangleq \thetak{\tr{t},k}{m} + \thetak{\tr{r},k}{n}$, where
\begin{IEEEeqnarray}{rCl}\label{eq:wiener_phn}
\thetak{\tr{t}, k}{m} &=& \thetak{\tr{t}, k-1}{m} + \Deltakn{\tr{t},k}{m}\nonumber\\
\thetak{\tr{r}, k}{n}&=&\thetak{\tr{r}, k-1}{n} + \Deltakn{\tr{r},k}{n}.
\end{IEEEeqnarray}
In \refeq{eq:wiener_phn}, $ \Deltakn{\tr{t},k}{m} \sim \mathcal{N}(0,\sigmap{\tr{t}}{2})$, $ \Deltakn{\tr{r},k}{n} \sim \mathcal{N}(0,\sigmap{\tr{r}}{2})$, and $\thetak{\tr{t},0}{m}$  and $\thetak{\tr{r},0}{n}$ are uniformly distributed in $[0, 2\pi)$. The phase noise in \refeq{eq:wiener_phn} is assumed to be varying from symbol-to-symbol, but constant on the support of the transmit pulse \cite{colavolpe05}.

Based on the received signal model in \refeq{eq:received_mimo} and \refeq{eq:wiener_phn}, we define the following vectors: $\bThetak{k} \triangleq [\thetak{\tr{t},k}{1}, \ldots, \thetak{\tr{t},k}{N_{\tr t}},  \thetak{\tr{r},k}{1}, \ldots, \thetak{\tr{r},k}{N_{\tr r}}]$, $\bbTheta \triangleq [\bThetak{1},\ldots,\bThetak{L} ]$, $\bck{k} \triangleq [\ck{k}{1}, \ldots, \ck{k}{N_{\tr t}}]$, and $\bbc  \triangleq [\bck{1},\ldots,\bck{L} ]$, $\brk{k} \triangleq [\rk{k}{1}, \ldots, \rk{k}{N_{\tr r}}]$, $\bbr \triangleq [\brk{1}, \ldots, \brk{L}]$, and $\bwk{k} \triangleq [\wk{k}{1}, \ldots, \wk{k}{N_{\tr r}}]$.

The following remarks are in order.
\begin{remark}
When the channel is unknown at the receiver, we assume that the channel and phase noise are jointly estimated as demonstrated in \cite{hani12}.
\end{remark}
\begin{remark}
The quality of the oscillators at the transmitter and the receiver depends on the values of $\sigmap{\tr{t}}{2}$ and $\sigmap{\tr{r}}{2}$, respectively. We implicitly assume that the quality of the oscillators at the transmit and the receive sides, respectively, are identical.
\end{remark}

\section{MAP Symbol Detector}
\label{sec:sbs_map}

In this section, we derive the MAP symbol detector. Based on the received signal model in \refeq{eq:received_mimo}, the optimum receiver is obtained as
\begin{IEEEeqnarray}{rCl}
\label{eq:sbs_map_a}
\hat{\bm c}_{k} &=& {\arg \underset{ \bck{k} } \max} \sum_{\bbc \setminus  \{\bck{k}\}}  P(\bbc|\bbr) \\
\label{eq:sbs_map_b}
&\propto& {\arg \underset{ \bck{k} } \max} \sum_{\bbc \setminus \{\bbc\}} P(\bck{k}) p(\bbr|\bbc)\nonumber\\
&=& {\arg \underset{ \bck{k} } \max} \; P(\bck{k})p(\bbr|\bck{k})\nonumber\\
&=& {\arg \underset{ \bck{k} } \max}  \underset{\bThetak{k}}{\int}  P(\bck{k})p(\bbr|\bck{k}, \bThetak{k}) p(\bThetak{k}|\bck{k})\tr d \bThetak{k}\nonumber\\
\label{eq:sbs_map_bb}
&=& {\arg \underset{ \bck{k} } \max}  \underset{\bThetak{k}}{\int}  P(\bck{k}) p(\brk{k}|\bck{k}, \bThetak{k},\bbr_{k}) p(\bbr_{k}|\bck{k}, \bThetak{k}) \nonumber\\
&& ~~~~~~~~~~~~~~\cdot p(\bThetak{k}|\bck{k})\tr d \bThetak{k} \\
\label{eq:sbs_map_c}
&=& {\arg \underset{ \bck{k} } \max}  \underset{\bThetak{k}}{\int}  P(\bck{k}) p(\brk{k}|\bck{k}, \bThetak{k}) p(\bThetak{k}|\bck{k}, \bbr_{k}) p( \bbr_{k}|\bck{k}) \tr d \bThetak{k}\nonumber\\
\\
\label{eq:sbs_map_d}
&\propto& {\arg \underset{ \bck{k} } \max}   \underset{\bThetak{k}}{\int}  P(\bck{k}) p(\brk{k}|\bck{k}, \bThetak{k}) p(\bThetak{k}| \bck{k}, \bbr_{k})  \tr d \bThetak{k}.
\end{IEEEeqnarray}
In \refeq{eq:sbs_map_a}, we express the MAP symbol detector for the symbols transmitted in the $k$th time instant as the marginalization of the a posteriori pmf of $\bbc$ with respect to all the symbols but $\bck{k}$. 
$P(\bck{k})$ represents the a priori probability of the transmitted symbols in the $k$th time instant. We define $\bbr_{k} \triangleq [\brk{1},\ldots,\brk{k-1},\brk{k+1},\ldots,\brk{L}]$ in \refeq{eq:sbs_map_bb}. In \refeq{eq:sbs_map_c}, it is applied that, given $\bck{k}$ and $\bThetak{k}$, $\bbr_{k}$ is independent of $\brk{k}$. It is assumed in \refeq{eq:sbs_map_d} that $\bck{k}$ and $\bbr_{k}$ are independent of each other, which is reasonable in the case of uncoded data transmission or in coded systems that employ a pseudo-random interleaver.

The detector obtained in \refeq{eq:sbs_map_d} is a vector extension of the MAP symbol detector derived by Kam \textit{et al.} in \cite{kam94} -- it detects $\bck{k}$ based on the conditional pdf of $\bThetak{k}$, $p(\bThetak{k}| \bck{k}, \bbr_{k})$, which is estimated using all received signals outside the $k$th time instant. The integral in \refeq{eq:sbs_map_d} represents the a posteriori pmf of the transmitted symbols that is obtained after the marginalization of the phase noise. In uncoded systems, the transmitted symbols are detected based on \refeq{eq:sbs_map_d}, while in coded systems, the a posteriori pmf of the transmitted symbols is used for computing the bit log-likelihood ratios (LLRs) for soft decoding \cite{rajet13}. For the system model in \refeq{eq:received_mimo}, computing the MAP symbol detector in \refeq{eq:sbs_map_d} is an infinite dimensional problem since $p(\bThetak{k}| \bck{k}, \bbr_{k})$ is a continuous function. This makes the MAP detector intractable \cite{kam94} and unimplementable in practice.

The MAP detector presented in \refeq{eq:sbs_map_d} can also be obtained by applying the SPA based on the factor graph framework \cite{loeliger01}. This analysis forms the basis of the algorithm that is presented in Section \ref{sec:spa}. In order to derive the MAP detector using the SPA, we rewrite \refeq{eq:sbs_map_a} as
\begin{IEEEeqnarray}{rCl}\label{eq:spa_sbsmap}
\hat{\bm c}_{k} &=& {\arg\underset{ \bck{k} } \max} \sum_{\bbc \setminus \{\bck{k}\}}  P(\bbc|\bbr) \nonumber\\
&=& {\arg\underset{ \bck{k} } \max}\sum_{\bbc \setminus \{\bck{k}\}} \underset{\bbTheta}{\int} P(\bbc,\bbTheta|\bbr)   \tr d \bbTheta,
\end{IEEEeqnarray}
Factorizing the integrand, we obtain
\begin{IEEEeqnarray}{rCl}\label{eq:spa_mapfactor}
P(\bbc,\bbTheta|\bbr) &\propto& P(\bbc)p(\bbTheta|\bbc)p(\bbr|\bbc,\bbTheta), \nonumber\\
&=& P(\bThetak{0})\prod_{k=1}^{L} P(\bck{k}) \underbrace{p(\bThetak{k}|\bThetak{k-1})}_{\Pdelta(\bThetak{k} - \bThetak{k-1})} p(\brk{k}|\bThetak{k},\bck{k}). \IEEEeqnarraynumspace
\end{IEEEeqnarray}
To factorize the function in \refeq{eq:spa_mapfactor} we exploit the fact that $\bThetak{k}$ is a discrete Wiener process as in \refeq{eq:wiener_phn}.

\begin{figure}[!t]
\begin{center}
\includegraphics[width = 3.6in, keepaspectratio=true]{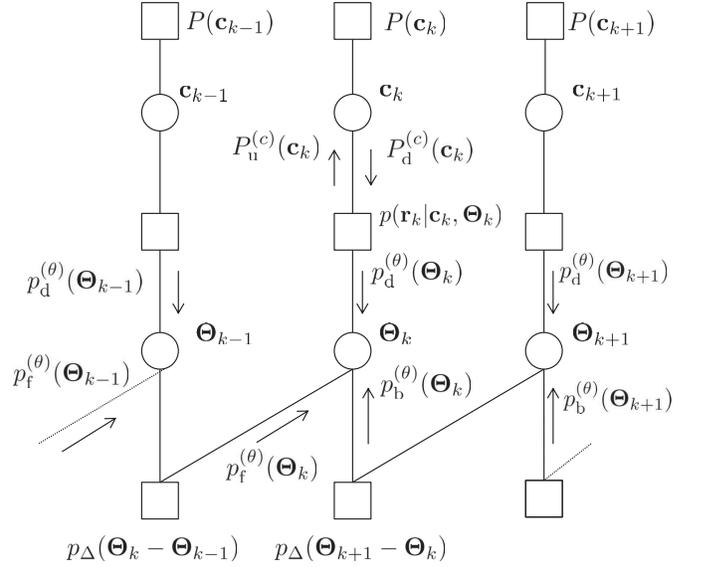}
\caption{ Factor Graph and the SPA messages based on \refeq{eq:spa_mapfactor}}
\label{fig:FGmimo}
\end{center}
\end{figure}

The factor graph (FG) associated with \refeq{eq:spa_mapfactor} is drawn in Fig.~\ref{fig:FGmimo}. With reference to the messages in the figure, we have
\begin{IEEEeqnarray}{rCl}
\label{eq:spa_msg_a}
\Pdc(\bck{k}) &=& P(\bck{k})\\
\Pdtheta(\bThetak{k}) &=& \sum_{\bck{k}} \Pdc(\bck{k})p(\brk{k} | \bck{k},\bThetak{k})  \\
\label{eq:spa_msg_b}
\Pftheta(\bThetak{k}) &=& \int_{\bThetak{k-1}} \Pftheta(\bThetak{k-1})\Pdtheta(\bThetak{k-1})\nonumber\\&& ~~~~~~~~\cdot\Pdelta(\bThetak{k} - \bThetak{k-1}) \tr d \bThetak{k-1}  \\
\label{eq:spa_msg_c}
\Pbtheta(\bThetak{k}) &=& \int_{\bThetak{k+1}} \Pbtheta(\bThetak{k+1})\Pdtheta(\bThetak{k+1})\nonumber\\&& ~~~~~~~~\cdot\Pdelta(\bThetak{k+1} - \bThetak{k}) \tr d \bThetak{k+1}   \\
\label{eq:spa_msg_d}
\Puc(\bck{k}) &=& \int_{\bThetak{k}}  \Pftheta(\bThetak{k})\Pbtheta(\bThetak{k})p(\brk{k} | \bck{k}, \bThetak{k} )\tr d \bThetak{k}.
\end{IEEEeqnarray}

Note that, in the case of uncoded transmission, the FG in Fig. \ref{fig:FGmimo} is a tree, and hence applying the SPA on this graph renders the exact MAP symbol detector \refeq{eq:sbs_map_d}. In this view, $\Pbtheta(\bThetak{k})\Pftheta(\bThetak{k})$ is equal to the a posteriori pdf $p(\bThetak{k}| \bck{k}, \bbr_{k})$ in \refeq{eq:sbs_map_d}. Thus, the detector in \refeq{eq:sbs_map_d} can be expressed in terms of $\Puc(\bck{k})$ as
\begin{IEEEeqnarray}{rCl} \label{eq:spa_map}
\hat{\bm c}_{k} &=& {\arg\underset{ \bck{k} }\max} \; \Puc(\bck{k}).
\end{IEEEeqnarray}

The messages in \refeq{eq:spa_msg_a}-\refeq{eq:spa_msg_d} form the core of the SPA for the implementation of the MAP detector. The implementation of the exact SPA is impractical because it involves the estimation of the continuous pdfs of $\bThetak{k}$ in \refeq{eq:spa_msg_a}-\refeq{eq:spa_msg_c} that entails infinite dimensionality. Hence the exact form of the messages are intractable. The intractability of the exact MAP symbol detector in \refeq{eq:sbs_map_d} and \refeq{eq:spa_map} motivates the need to explore practical, low complexity receiver algorithms, which are investigated in the sequel.

\section{Multivariate Tikhonov-Parameterization Based Sum-Product Algorithm for Approximate MAP Detection}
\label{sec:spa}

In the following, we derive a low-complexity SPA for the approximate implementation of the MAP symbol detector based on the canonical distribution approach suggested in \cite{worthen01}.  This approach involves constraining the messages on the FG to a specific family of pdfs that can compactly and completely be described by a finite number of parameters. Thus, the task of computing the exact pdf is reduced to computing the parameters of the pdf. More specifically, we adopt the Tikhonov canonical distribution approach introduced by Colavolpe \textit{et al.} in \cite{colavolpe05}; we constrain $\Pftheta(\bThetak{k})$ and  $\Pbtheta(\bThetak{k})$ to be multivariate Tikhonov pdfs in order to obtain a practical algorithm with good performance.

Without loss of generality we consider the case where $N_{\tr t} = 2$ and $N_{\tr r} = 1$, hence $\bThetak{k} = [\thetak{\tr{t},k}{1}, \thetak{\tr{t},k}{2}, \thetak{\tr{r},k}{1}]$ and $\bck{k} = [\ck{k}{1}, \ck{k}{2}]$. The generalization of the algorithm to arbitrary values of $N_{\tr t}$ and $N_{\tr r}$ is straightforward and is presented in Section \ref{sec:spa_gen}. The received signal model in the $k$th time instant is
\begin{IEEEeqnarray}{rCl}\label{eq:sys_mod2x1}
\rk{k}{1} =  \ck{k}{1} e^{j (\thetak{\tr{t},k}{1} + \thetak{\tr{r},k}{1})} + \ck{k}{2} e^{j (\thetak{\tr{t},k}{2} + \thetak{\tr{r},k}{1})} + \wk{k}{1}.
\end{IEEEeqnarray}
Then \refeq{eq:spa_mapfactor} can be expressed as
\begin{IEEEeqnarray}{rCl}\label{eq:spa_factor2x1}
P(\bbc,\bbTheta|\bbr) &\propto&   P(\bbc) p(\thetak{\tr t,0}{1}, \thetak{\tr t,0}{2},\thetak{\tr r,0}{1}) \prod_{k}\Pdelta (\bThetak{k} - \bThetak{k-1}) \nonumber\\ &&\prod_{k} p(\rk{k}{1}|\ck{k}{1},\ck{k}{2},\thetak{\tr t,k}{ 1},\thetak{\tr t,k}{ 2},\thetak{\tr r, k}{1}),\label{eq:pdf}
\end{IEEEeqnarray}
where
\begin{IEEEeqnarray}{rCl}\label{eq:r_ctheta2x1}
 && p(\rk{k}{1}|\ck{k}{1},\ck{k}{2},\thetak{\tr t,k}{1},\thetak{\tr t,k}{ 2},\thetak{\tr r, k}{1})
 \nonumber \\ &&\propto   \exp\!\left\{ -\frac{\left|\rk{k}{1}-\ck{k}{1}e^{\jmath (\thetak{\tr{t},k}{1}  + \thetak{\tr{r},k}{1})}-\ck{k}{2}e^{\jmath (\thetak{\tr{t},k}{2}  + \thetak{\tr{r},k}{1})}\right|^{2}}{N_{0}}\right\}. \IEEEeqnarraynumspace
\end{IEEEeqnarray}
We first seek to determine the functional form of the message $\Pdtheta(\bThetak{k})$ which is used to determine the messages $\Pftheta(\bThetak{k})$ and $\Pbtheta(\bThetak{k})$. From \refeq{eq:spa_msg_a},
\begin{IEEEeqnarray}{rcl}\label{eq:spa_pdtheta}
\Pdtheta(\bThetak{k})&=&\sum_{\ck{k}{1}}\sum_{\ck{k}{2}}\Pdc(\ck{k}{1},\ck{k}{2})p(\rk{k}{1}|\ck{k}{1},\ck{k}{2},\thetak{\tr t,k}{1},\thetak{\tr t,k}{ 2},\thetak{\tr r, k}{1})\nonumber\\
&=& p(r_{k}|\thetak{\tr t,k}{1},\thetak{\tr t,k}{ 2},\thetak{\tr r, k}{1}).
\end{IEEEeqnarray}
We approximate $p(r_{k}|\thetak{\tr t,k}{1},\thetak{\tr t,k}{ 2},\thetak{\tr r, k}{1})$ by the Gaussian pdf that is closest in terms of the Kullbach Leibler (KL) divergence measure. This is achieved by moment matching, since the Gaussian pdf belongs to the exponential family of distributions \cite{minka05}. The mean and variance of the closest Gaussian pdf are
\begin{IEEEeqnarray}{rCl}\label{eq:pdtheta_moment_match}
\mathbb E \{\rk{k}{1}|\thetak{\tr t,k}{1},\thetak{\tr t,k}{ 2},\thetak{\tr r, k}{1}\} & = & \alphak{k}{1}e^{\jmath (\thetak{\tr{t},k}{1} + \thetak{\tr{r},k}{1})} +\alphak{k}{2}e^{\jmath (\thetak{\tr{t},k}{2} + \thetak{\tr{r},k}{1})} \nonumber\\
{\mathrm{Var}}\{\rk{k}{1}|\thetak{\tr t,k}{1},\thetak{\tr t,k}{ 2},\thetak{\tr r, k}{1}\} & = & \betak{k}{1}+\betak{k}{2} + N_{0}- \left|\alphak{k}{1}\right|^{2}-\left|\alphak{k}{2}\right|^{2} \nonumber\\
&\triangleq& \gamma_{k},
\end{IEEEeqnarray}
respectively, having defined
\begin{IEEEeqnarray}{rCl} \label{eq:alphabeta}
\alphak{k}{i} & = & \sum_{\ck{k}{i} \in \C}\ck{k}{i}\Pdc(\ck{k}{i})\\
\betak{k}{i} & = & \sum_{\ck{k}{i} \in \C}\left|\ck{k}{i}\right|^{2}\Pdc(\ck{k}{i}), \text{for } i = \{1,\ldots,N_{\tr t}\}.
\end{IEEEeqnarray}
Therefore,
\begin{IEEEeqnarray}{rCl}\label{eq:pdtheta}
&&\Pdtheta(\bThetak{k}) \nonumber\\
&&\approx \N(\rk{k}{1}; \mathbb E \{\rk{k}{1}|\thetak{\tr t,k}{1},\thetak{\tr t,k}{ 2},\thetak{\tr r, k}{1}\}, \gamma_{k}) \\
&&\propto  \exp\!\left\{ -\frac{\left|\rk{k}{1}-\alphak{k}{1}e^{\jmath (\thetak{\tr{t},k}{1}  + \thetak{\tr{r},k}{1})}-\alphak{k}{2}e^{\jmath (\thetak{\tr{t},k}{2}  + \thetak{\tr{r},k}{1})}\right|^{2}}{\gamma_{k}}\right\} \nonumber\\
&&\propto \exp\!\left\{ \frac{2}{\gamma_{k}}\Re\left[\rk{k}{1}{\alphak{k}{1}}^{*}e^{-\jmath (\thetak{\tr{t},k}{1}  + \thetak{\tr{r},k}{1})} + \rk{k}{1}{\alphak{k}{2}}^{*}e^{-\jmath (\thetak{\tr{t},k}{2}  + \thetak{\tr{r},k}{1})} \right.\right. \nonumber\\ &&\left.\left. ~~~~~~~~-~ |\alphak{k}{2}{\alphak{k}{1}}^{*}|e^{\jmath( \angle{\alphak{k}{2}{\alphak{k}{1}}^{*}} +  \thetak{\tr{t},k}{2}-\thetak{\tr{t},k}{1})}\right]\right\} \nonumber\\
\label{eq:pdtheta_a}
&&= \exp\!\left\{ \frac{2}{\gamma_{k}}\Re\left[\rk{k}{1}{\alphak{k}{1}}^{*}e^{-\jmath (\thetak{\tr{t},k}{1}  + \thetak{\tr{r},k}{1})} + \rk{k}{1}{\alphak{k}{2}}^{*}e^{-\jmath (\thetak{\tr{t},k}{2}  + \thetak{\tr{r},k}{1})}  \right.\right. \nonumber\\ &&\left.\left. ~~~~~~~~-~ |\alphak{k}{2}{\alphak{k}{1}}^{*}|e^{\jmath( \angle{\rk{k}{1}{\alphak{k}{1}}^{*}} - \angle{\rk{k}{1}{\alphak{k}{2}}^{*}} +  \thetak{\tr{t},k}{2}-\thetak{\tr{t},k}{1})}\right]\right\} \\
\label{eq:pdtheta_b}
&&\triangleq \exp\!\left\{ \Re\left[(\xk{k}{1}e^{-\jmath \thetak{\tr{t},k}{1}}+ \xk{k}{2}e^{-\jmath \thetak{\tr{t},k}{2}})e^{-\jmath \thetak{\tr{r},k}{1}} \right.\right. \nonumber\\ && \left.\left.  ~~~~~~~~-~\xk{k}{3}  e^{-\jmath(\thetak{\tr{t},k}{1}-\thetak{\tr{t},k}{2})}\right]\right\} \IEEEeqnarraynumspace\\
\label{eq:pdtheta_c}
&&= \exp\!\left\{ \Re\left[\xk{k}{1}e^{-\jmath \thetak{k}{1,1}}+ \xk{k}{2}e^{-\jmath \thetak{k}{2,1}}  \right.\right. \nonumber\\ && \left.\left.  ~~~~~~~~-~ \xk{k}{3}e^{-\jmath(\thetak{k}{1,1}-\thetak{k}{2,1})}\right]\right\}.\IEEEeqnarraynumspace
\end{IEEEeqnarray}
In \refeq{eq:pdtheta_a}, we exploit that $\angle{\alphak{k}{2}{\alphak{k}{1}}^{*}} = \angle{\rk{k}{1}{\alphak{k}{1}}^{*}} - \angle{\rk{k}{1}{\alphak{k}{2}}^{*}}$, and in \refeq{eq:pdtheta_b} we define
\begin{IEEEeqnarray}{rCl}\label{eq:bi_tikhnv}
\xk{k}{1}  &\triangleq&  \frac{2}{\gamma_{k}}\left|\rk{k}{1}{\alphak{k}{1}}^{*}\right|e^{\jmath \angle \rk{k}{1}{\alphak{k}{1}}^{*} } \nonumber\\
\xk{k}{2}  &\triangleq&  \frac{2}{\gamma_{k}}\left|\rk{k}{1}{\alphak{k}{2}}^{*}\right|e^{\jmath  \angle \rk{k}{1}{\alphak{k}{2}}^{*} }\nonumber\\
\xk{k}{3}  &\triangleq&  \frac{2}{\gamma_{k}}\left|\alphak{k}{2}{\alphak{k}{1}}^{*}\right|e^{\jmath  (\angle \rk{k}{1}{\alphak{k}{1}}^{*} - \angle \rk{k}{1}{\alphak{k}{2}}^{*} )}.
\end{IEEEeqnarray}
In \refeq{eq:pdtheta_c}, the message $\Pdtheta(\bThetak{k})$ is rewritten as the cosine variant of the unnormalized bivariate Tikhonov distribution of $\thetak{k}{1,1}, \thetak{k}{2,1} $ \cite{mardia07}. The distribution is completely characterized by $\xk{k}{1},\xk{k}{2}, \xk{k}{3}$ and its parameters are the following. $\angle{\xk{k}{1}}$ and  $1/|{\xk{k}{1}}|$ are the mean and variance of $\thetak{k}{1,1}$, respectively, $\angle{\xk{k}{2}}$ and  $1/|{\xk{k}{2}}|$ are the mean and variance of $\thetak{k}{2,1}$, respectively, and $\xk{k}{3}$ is related to the correlation between $\thetak{k}{1,1}$ and $\thetak{k}{2,1}$, which can have an arbitrary magnitude and has to satisfy the constraint $\angle \xk{k}{3} =\angle \xk{k}{1} - \angle \xk{k}{2}$. The estimates of the states $\thetak{k}{1,1},\thetak{k}{2,1}$ and their covariance mentioned are based on the received signal $\rk{k}{1}$. In our algorithm, we exploit the functional form of $\Pdtheta(\bThetak{k})$ in \refeq{eq:pdtheta_b} to determine the other messages.

\subsection{Forward Recursion}
In the sequel, based on $\Pdtheta(\bThetak{k})$, we determine the message $\Pftheta(\bThetak{k})$, which is constrained to be a bivariate Tikhonov pdf. Computation of the parameters of $\Pftheta(\bThetak{k})$ renders the state estimates and their covariance based on the received signals $[\rk{1}{1},\ldots,\rk{L}{1}]$ in the forward direction -- this is referred to as the forward recursion. The message is evaluated as
\begin{IEEEeqnarray}{rCl}\label{eq:spa_pftheta}
\Pftheta(\bThetak{k}) = \underset{\bThetak{k}}\int \Pftheta(\bThetak{k-1})\Pdtheta(\bThetak{k-1})\Pdelta(\bThetak{k}-\bThetak{k-1})\tr d\bThetak{k-1}.\nonumber\\
\end{IEEEeqnarray}
Assume that $\Pftheta(\bThetak{k-1})$ is the cosine variant of the bivariate Tikhonov distribution and is given as
\begin{IEEEeqnarray}{rCl} \label{eq:pftheta_km1}
\Pftheta(\bThetak{k-1})
&\propto&\exp\!\left\{ \Re\left[(\afk{k-1}{1,1}e^{-\jmath \thetak{\tr{t},k-1}{1}}+\afk{k-1}{2,1}e^{-\jmath\thetak{\tr{t}, k-1}{2}}) \right.\right. \nonumber\\ &&\left.\left. ~~~~~~~ \cdot e^{-\jmath\thetak{\tr{r}, k-1}{1}} - \tafk{k-1}{1,2}e^{-\jmath(\thetak{\tr{t}, k-1}{1} - \thetak{\tr{t}, k-1}{2})}\right]\right\},  \nonumber\\
\end{IEEEeqnarray}
where $\angle \tafk{k-1}{1,2} = \angle \afk{k-1}{1,1}-\angle \afk{k-1}{2,1}$. In \refeq{eq:pftheta_km1}, $\angle \afk{k-1}{1}$, $\angle \afk{k-1}{2}$ and $1/|\afk{k-1}{1}|$, $1/|\afk{k-1}{2}|$ correspond to the predicted state estimates and the variances of $\thetak{k-1}{1,1}$, $\thetak{k-1}{2,1}$, respectively, based on the received signals $[\rk{1}{1},\ldots,\rk{k-1}{1}]$, and $\tafk{k-1}{1,2}$ gives a measure of the predicted correlation between the states.
Now compute the product $\Pftheta(\bThetak{k-1})\Pdtheta(\bThetak{k-1})$ in \refeq{eq:spa_pftheta} as
\begin{IEEEeqnarray}{rCl}\label{eq:pfpd}
&& \Pftheta(\bThetak{k-1})\Pdtheta(\bThetak{k-1}) \nonumber\\
&&=\! \exp\!\left\{ \Re\left[ ((\xk{k-1}{1} + \afk{k-1}{1,1})e^{-\jmath \thetak{\tr{t},k-1}{1}} + (\xk{k-1}{2}+\afk{k-1}{2,1})e^{-\jmath \thetak{\tr{t},k-1}{2}}) \right.\right. \nonumber\\
&& \left.\left. ~~~~~~~ \cdot e^{-\jmath\thetak{\tr{r},k-1}{1}} - (\xk{k-1}{3} + \tafk{k-1}{1,2})e^{-\jmath(\thetak{\tr{t},k-1}{1}-\thetak{\tr{t},k-1}{2})}\right]\right\} \nonumber\\
\label{eq:prod}
&&\triangleq \exp\!\left\{ \Re\left[(\yk{k-1}{1}e^{-\jmath \thetak{\tr{t},k-1}{1}}+\yk{k-1}{2}e^{-\jmath\thetak{\tr{t}, k-1}{2}})e^{-\jmath\thetak{\tr{r}, k-1}{1}} \right.\right. \nonumber\\
 &&\left.\left. ~~~~~~-~\yk{k-1}{3}e^{-\jmath(\thetak{\tr{t}, k-1}{1} - \thetak{\tr{t}, k-1}{2})}\right]\right\}.
\end{IEEEeqnarray}

In \refeq{eq:prod}, $\angle \yk{k-1}{1}$, $\angle \yk{k-1}{2}$ and $1/|\yk{k-1}{1}|$, $1/|\yk{k-1}{2}|$ correspond to the predicted state estimates of $\thetak{k-1}{1,1}$, $\thetak{k-1}{2,1}$, respectively, and their variances based on the received signals $[\rk{1}{1},\ldots,\rk{k}{1}]$, and $\yk{k-1}{3}$ gives a measure of the predicted correlation between the states.

Note that bivariate Tikhonov distributions are not closed under the product operation \cite{charles09}, i.e., the product of two bivariate Tikhonov distributions is not another bivariate Tikhonov distribution. However, as in \cite{charles09}, we consider that the product in \refeq{eq:pfpd} can be approximated as a bivariate Tikhonov distribution, where we assume that  $\angle \yk{k-1}{3}\approx \angle \yk{k-1}{1} - \angle \yk{k-1}{2}$. Using this assumption, we compute $\Pftheta(\bThetak{k})$ in \refeq{eq:spa_pftheta} as
\begin{IEEEeqnarray}{rCl}\label{eq:pfthetak}
&&\Pftheta(\bThetak{k}) \nonumber\\ &&= \int_{0}^{2\pi}\int_{0}^{2\pi}\int_{0}^{2\pi}  \exp\!\left\{ \Re\left[(\yk{k-1}{1}e^{-\jmath \thetak{\tr{t},k-1}{1}}+\yk{k-1}{2}e^{-\jmath\thetak{\tr{t}, k-1}{2}})\right.\right. \nonumber\\
&& \left.\left. ~~~~~~~~~~~~~~~~~~~~~~~~~~\cdot e^{-\jmath\thetak{\tr{r}, k-1}{1}} - \yk{k-1}{3}e^{-\jmath(\thetak{\tr{t}, k-1}{1} - \thetak{\tr{t}, k-1}{2})}\right]\right\} \nonumber\\
&& ~~~~~~~~~~~~~~~~~~~~ \cdot \Pdelta(\thetak{\tr{t}, k}{1} - \thetak{\tr{t}, k-1}{1})\Pdelta(\thetak{\tr{t}, k}{2} - \thetak{\tr{t}, k-1}{2}) \nonumber\\
&& ~~~~~~~~~~~~~~~~~~~~~ \cdot \Pdelta(\thetak{\tr{r}, k}{1} - \thetak{\tr{r}, k-1}{1})  \tr d \thetak{\tr{t}, k-1}{1}  \tr d \thetak{\tr{t}, k-1}{2} \tr d \thetak{\tr{r}, k-1}{1}.\IEEEeqnarraynumspace
\end{IEEEeqnarray}

For the discrete Wiener phase noise process considered, we show in Appendix A that $\Pftheta(\bThetak{k})$ is approximately a bivariate Tikhonov distribution given by
\begin{IEEEeqnarray}{rCl}\label{eq:pfthetak_final}
\Pftheta(\bThetak{k}) & \propto & \exp\!\left\{ \Re\left[(\afk{k}{1,1}e^{-\jmath \thetak{\tr{t},k}{1}}+\afk{k}{2,1}e^{-\jmath\thetak{\tr{t}, k}{2}})e^{-\jmath\thetak{\tr{r}, k}{1}} \right.\right. \nonumber\\ &&\left.\left. ~~~~~-~\tafk{k}{1,2}e^{-\jmath(\thetak{\tr{t}, k}{1} - \thetak{\tr{t}, k}{2})}\right]\right\},
\end{IEEEeqnarray}
where it is assumed that $\angle \tafk{k}{1,2} = \angle \afk{k}{1,1}-\angle \afk{k}{2,1}$. The parameters $\afk{k}{1,1}, \afk{k}{2,1}, \tafk{k}{1,2}$ are recursively updated in the forward direction as
\begin{IEEEeqnarray}{rCl}\label{eq:afkupdate_fstphn}
\afk{k}{m,1}& = & \frac{\bbafk{k}{m,1}}{1+\sigmap{\tr{t}}{2}\left|\left|\bbafk{k}{m,1}\right|-\left|\ttafk{k}{1,2}\right|\right|},~\mbox{$m \in \{1,2\}$} \nonumber\\
\tafk{k}{1,2}  & = & \frac{\ttafk{k}{1,2}}{\overset{2}{\underset{m=1}{\prod}}\left({1+\sigmap{\tr{t}}{2}\left|\left|\bbafk{k}{m,1}\right|-\left|\ttafk{k}{1,2}\right|\right|}\right)},
\end{IEEEeqnarray}
where
\begin{IEEEeqnarray}{rCl}\label{eq:afkupdate_fstphn1}
\bbafk{k}{m,1} & = & \frac{\bafk{k}{m,1}}{1+\sigmap{\tr{r}}{2}\left|\left|\bafk{k}{1,1}\right|+\left|\bafk{k}{2,1}\right|\right|}.\nonumber\\
\bafk{k}{m,1} & = & \afk{k-1}{m,1}+\frac{2}{\gamma_{k-1}}\rk{k-1}{1}{\alphak{k-1}{m}}^{*} \nonumber\\
\ttafk{k}{1,2} & = & \tafk{k-1}{1,2}+\frac{2}{\gamma_{k-1}}\alphak{k-1}{2}{\alphak{k-1}{1}}^{*}
\end{IEEEeqnarray}

\subsection{Backward Recursion}

The parameters of $\Pbtheta(\bThetak{k})$ are computed based on the received signals $[\rk{L}{1},\ldots,\rk{1}{1}]$ in the backward direction. The message $\Pbtheta(\bThetak{k})$ in \refeq{eq:spa_msg_c} is evaluated as
\begin{IEEEeqnarray}{rCl}\label{eq:pbthetak_final}
\Pbtheta(\bThetak{k}) & \propto & \exp\!\left\{ \Re\left[(\abk{k}{1,1}e^{-\jmath \thetak{\tr{t},k}{1}}+\abk{k}{2,1}e^{-\jmath\thetak{\tr{t}, k}{2}})e^{-\jmath\thetak{\tr{r}, k}{1}} \right.\right. \nonumber\\ &&\left.\left. ~~~~~-~\tabk{k}{1,2}e^{-\jmath(\thetak{\tr{t}, k}{1} - \thetak{\tr{t}, k}{2})}\right]\right\},
\end{IEEEeqnarray}
where it is assumed that $\angle \tabk{k}{1,2} = \angle \abk{k}{1,1} -\angle \abk{k}{2,1}$. The parameters of $\Pbtheta(\bThetak{k})$ are recursively updated in the backward direction as
\begin{IEEEeqnarray}{rCl}\label{eq:abkupdate_fstphn}
\abk{k}{m,1}& = & \frac{\bbabk{k}{m,1}}{1+\sigmap{\tr{t}}{2}\left|\left|\bbabk{k}{m,1}\right|-\left|\ttabk{k}{1,2}\right|\right|},~\mbox{$m \in \{1,2\}$} \nonumber\\
\tabk{k}{1,2}  & = & \frac{\ttabk{k}{1,2}}{\overset{2}{\underset{m=1}{\prod}}\left({1+\sigmap{\tr{t}}{2}\left|\left|\bbabk{k}{m,1}\right|-\left|\ttabk{k}{1,2}\right|\right|}\right)},
\end{IEEEeqnarray}
where
\begin{IEEEeqnarray}{rCl}\label{eq:abkupdate_fstphn1}
\bbabk{k}{m,1} & = & \frac{\babk{k}{m,1}}{1+\sigmap{\tr{r}}{2}\left|\left|\babk{k}{1,1}\right|+\left|\babk{k}{2,1}\right|\right|}\nonumber\\
\babk{k}{m,1} & = & \abk{k+1}{m,1}+\frac{2}{\gamma_{k+1}}\rk{k+1}{1}{\alphak{k+1}{m}}^{*} \nonumber\\
\ttabk{k}{1,2} & = & \tabk{k+1}{1,2}+\frac{2}{\gamma_{k+1}}\alphak{k+1}{2}{\alphak{k+1}{1}}^{*}.
\end{IEEEeqnarray}

\subsection{Computation of $\Puc(\bck{k})$}
Based on the messages $\Pftheta(\bThetak{k})$ and $\Pbtheta(\bThetak{k})$, we compute $\Puc(\bck{k})$ in \refeq{eq:spa_msg_d} as
\begin{IEEEeqnarray}{rCl}\label{eq:puck_fstphn}
\Puc(\bck{k}) & =&  \underset{\bThetak{k}}\int \Pftheta(\bThetak{k})\Pbtheta(\bThetak{k})p(\brk{k}|\bck{k},\bThetak{k}) \tr d \bThetak{k} \nonumber\\
&\propto&  \exp\!{\left(-\frac{|\ck{k}{1}|^{2}+|\ck{k}{2}|^{2}}{N_{0}}\right)} \nonumber\\ && \cdot \underset{\bThetak{k}}\int \exp\!\left\{ \Re\left[(\zk{k}{1,1}e^{-\jmath \thetak{\tr{t},k}{1}} +\zk{k}{2,1}e^{-\jmath\thetak{\tr{t}, k}{2}})e^{-\jmath\thetak{\tr{r}, k}{1}} \right.\right. \nonumber\\ && \left.\left. ~~~~~~~~-~\tzk{k}{1,2}e^{-\jmath(\thetak{\tr{t}, k}{1} - \thetak{\tr{t}, k}{2})}\right]\right\} \tr d \bThetak{k} \nonumber\\
&\propto&\exp\!{\left(-\frac{|\ck{k}{1}|^{2}+|\ck{k}{2}|^{2}}{N_{0}}\right)} \tr I_{0} \left( \left|\zk{k}{1,1}\right| +\left|\zk{k}{2,1}\right|\right) \nonumber\\
&& ~\cdot \tr I_{0}\left(\left|\tzk{k}{1,2}\right|\right),
\end{IEEEeqnarray}
where $ \tr I_{0}(\cdot)$ is the zeroth order modified Bessel function, and we define
\begin{IEEEeqnarray}{rCl}\label{eq:z_puck}
\zk{k}{m,1} & \triangleq & \afk{k}{m,1}+\abk{k}{m,1}+\frac{2}{N_{0}}\rk{k}{1}{\ck{k}{m}}^{*},~{m \in \{1,2\}}\nonumber\\
\tzk{k}{1,2} & \triangleq & \tafk{k}{1,2}+\tabk{k}{1,2}+\frac{2}{N_{0}}\ck{k}{2}{\ck{k}{1}}^{*}.
\end{IEEEeqnarray}

\subsection{Generalization to Arbitrary $N_{\tr t}$ and $N_{\tr r}$ values }
\label{sec:spa_gen}

Based on \refeq{eq:afkupdate_fstphn}, we can generalize the forward recursions for the case of arbitrary $N_{\tr t}$ and $N_{\tr r}$ values as
\begin{IEEEeqnarray}{rCl}\label{eq:afkupdate_gen}
\afk{k}{m,n} & = & \frac{ \underset{n=1}{\overset{N_{\tr r}}{\sum}}  \bbafk{k}{m,n}}{1+\sigmap{\tr{t}}{2}\left| \underset{n=1}{\overset{N_{\tr r}}{\sum}} \left|\bbafk{k}{m,n}\right|-\underset{\underset{l\neq m}{l=1}}{\overset{N_{\tr t}}{\sum}}\left|\ttafk{k}{m,l}\right|\right|}\nonumber\\
\tafk{k}{m,l}  & = & \frac{\ttafk{k}{m,l}}{\underset{m=1}{\overset{N_{\tr t}}{\prod}}\left({1+\sigmap{\tr{t}}{2}\left|\underset{n=1}{\overset{N_{\tr r}}{\sum}} \left|\bbafk{k}{m,n}\right|- \underset{\underset{l\neq m}{l=1}}{\overset{N_{\tr t}}{\sum}}\left|\ttafk{k}{m,l}\right|\right|}\right)},
\end{IEEEeqnarray}
where $m, l \in \{1,\ldots,N_{\tr t} \}$ and $n \in \{1,\ldots,N_{\tr r} \}$ and
\begin{IEEEeqnarray}{rCl}\label{eq:afkupdate_gen1}
\bbafk{k}{m,n} & = & \frac{\bafk{k}{m,n}}{1+\sigmap{\tr{r}}{2} \underset{m=1}{\overset{N_{\tr t}}{\sum}}\left|\bafk{k}{m,n}\right|}  \nonumber\\
\bafk{k}{m,n} & = & \afk{k-1}{m,n}+\frac{2}{\gamma_{k-1}}\rk{k-1}{n}{\alphak{k-1}{m}}^{*}  \nonumber\\
\ttafk{k}{m,l} & = &  \underset{n=1}{\overset{N_{\tr r}}{\sum}}\left( \tafk{k-1}{m,l}+\frac{2}{\gamma_{k-1}}\alphak{k-1}{l}{\alphak{k-1}{m}}^{*} \right).
\end{IEEEeqnarray}

Similarly, we can generalize the backward recursion as
\begin{IEEEeqnarray}{rCl}\label{eq:abkupdate_gen}
\abk{k}{m,n} & = & \frac{ \underset{n=1}{\overset{N_{\tr r}}{\sum}}  \bbabk{k}{m,n}}{1+\sigmap{\tr{t}}{2}\left| \underset{n=1}{\overset{N_{\tr r}}{\sum}} \left|\bbabk{k}{m,n}\right|-\underset{\underset{l\neq m}{l=1}}{\overset{N_{\tr t}}{\sum}}\left|\ttabk{k}{m,l}\right|\right|}\nonumber\\
\tabk{k}{m,l}  & = & \frac{\ttabk{k}{m,l}}{\underset{m=1}{\overset{N_{\tr t}}{\prod}}\left({1+\sigmap{\tr{t}}{2}\left|\underset{n=1}{\overset{N_{\tr r}}{\sum}} \left|\bbabk{k}{m,n}\right|- \underset{\underset{l\neq m}{l=1}}{\overset{N_{\tr t}}{\sum}}\left|\ttabk{k}{m,l}\right|\right|}\right)},
\end{IEEEeqnarray}
where
\begin{IEEEeqnarray}{rCl}\label{eq:abkupdate_gen1}
\bbabk{k}{m,n} & = & \frac{\babk{k}{m,n}}{1+\sigmap{\tr{r}}{2} \underset{m=1}{\overset{N_{\tr t}}{\sum}}\left|\babk{k}{m,n}\right|}  \nonumber\\
\babk{k}{m,n} & = & \abk{k+1}{m,n}+\frac{2}{\gamma_{k+1}}\rk{k+1}{n}{\alphak{k+1}{m}}^{*}  \nonumber\\
\ttabk{k}{m,l} & = &  \underset{n=1}{\overset{N_{\tr r}}{\sum}} \left(\tafk{k+1}{m,l}+\frac{2}{\gamma_{k+1}}\alphak{k+1}{l}{\alphak{k+1}{m}}^{*} \right).
\end{IEEEeqnarray}
The generalization of $\Puc(\bck{k})$ is given as
\begin{IEEEeqnarray}{rCl}\label{eq:puck_fstphn_gen}
\Puc(\bck{k}) &\propto&  \exp\!{\left(-N_{\tr r}\underset{m=1}{\overset{N_{\tr t}}{\sum}}\frac{|\ck{k}{m}|^{2}}{N_{0}}\right)} \underset{n=1}{\overset{N_{\tr r}}{\prod}} I_{0} \left(\underset{m=1}{\overset{N_{\tr t}}{\sum}} \left|\zk{k}{m,n}\right|\right) \nonumber\\ &&  \underset{\stackrel{m=1,}{l > m}}{\overset{N_{\tr t}}{\prod}} \tr I_{0} \left( \left|\tzk{k}{m,l}\right|\right),
\end{IEEEeqnarray}
where
\begin{IEEEeqnarray}{rCl}\label{eq:z_puck_gen}
\zk{k}{m,n} & \triangleq & \afk{k}{m,n}+\abk{k}{m,n}+\underset{n=1}{\overset{N_{\tr r}}{\sum}}  \frac{2}{N_{0}}\rk{k}{n}{\ck{k}{m}}^{*} \nonumber\\
\tzk{k}{m,l} & \triangleq & \tafk{k}{m,l}+\tabk{k}{m,l}+ \underset{n=1}{\overset{N_{\tr r}}{\sum}} \frac{2}{N_{0}}\ck{k}{l}{\ck{k}{m}}^{*}.
\end{IEEEeqnarray}

For future reference, we refer to the SPA-based algorithm for approximate MAP detection as SPA-MAP. We summarize one iteration of the SPA-MAP algorithm as follows.
\begin{itemize} [align=left,style=nextline,leftmargin=*,labelsep=\parindent,font=\normalfont]
    \item[ Step $1$)] Evaluate the coefficients $\alphak{k}{i}, \betak{k}{i}$ in \refeq{eq:alphabeta} using the a priori probabilities $\Pdc(\bck{k})$ of the transmitted symbols $\ck{k}{i} \forall\, k \in \{1,\ldots,L\} $ and $i \in \{1,\ldots,N_{\tr t}\} $.
    \item[ Step $2$)] Recursively update the parameters in \refeq{eq:afkupdate_gen} in the forward direction using the received signals $[\rk{1}{j},\ldots,\rk{L}{j}] \forall\, j \in \{1,\ldots,N_{\tr r}\} $.
    \item[ Step $3$)] Recursively update the parameters in \refeq{eq:abkupdate_gen} in the backward direction using the received signals $[\rk{L}{j},\ldots,\rk{1}{j}] \forall\, j \in \{1,\ldots,N_{\tr r}\} $.
    \item[ Step $4$)] Evaluate $\Puc(\bck{k})$ in \refeq{eq:puck_fstphn_gen} $\forall\, k \in \{1,\ldots,L\}$.
    \item[ Step $5$)] Update $\Pdc(\bck{k})$ using $\Puc(\bck{k})$, if a soft-input soft-output decoder is used $\forall\, k \in \{1,\ldots,L\} $.
\end{itemize}

\section{Approximate MAP Detection Based on the Smoother-Detector Structure}
\label{sec:estdet}

In this section, we present a receiver algorithm that uses the smoother-detector structure, as in \cite{rajet13}, to approximate the MAP symbol detector. As required by this structure, a smoother like the EKS is used to track the discrete Wiener phase noise process. Then the a posteriori pdf from the smoother is used for deriving the approximate MAP detector in \refeq{eq:sbs_map_d}. Specifically, let  $p(\bThetak{k}| \bbr)$ denote the a posteriori phase noise pdf provided by the smoother. We use this pdf in \refeq{eq:sbs_map_d} by assuming that $p(\bThetak{k}| \bck{k}, \bbr_{k}) \approx p(\bThetak{k}| \bbr)$, and $p(\bThetak{k}| \bbr)$ is further approximated by constraining it to a specific family of pdfs that renders the integral in \refeq{eq:sbs_map_d} tractable.

We first consider the case where $N_{\tr t} = 2$ and $N_{\tr r} = 1$, and, with a slight abuse of notation, we let $\bThetak{k} = [\thetak{k}{1,1}, \thetak{k}{2,1}]$ and $\bck{k} = [\ck{k}{1}, \ck{k}{2}]$. The pdf $p(\bThetak{k}| \bbr)$ is modeled as a bivariate Gaussian pdf, i.e., $p(\bThetak{k}| \bbr) = \N(\bThetak{k};\bThetahatk{k},\mathbf{{P}}_{k,n})$, where $\thetahatk{k}{m,n} \in \bThetahatk{k}$ and $P_{k,n}^{(m,l)} \in \mathbf{P}_{k,n}$, with
\begin{IEEEeqnarray}{rCl}\label{eq:mean_var_est}
&&\thetahatk{k}{m,n} = \mathcal{\mathbb{E}}_{p(\bThetak{k}| \bbr)}\{\thetak{k}{m,n}\},\nonumber\\
&&P_{k,n}^{(m,l)} = \mathcal{\mathbb{E}}_{p(\bThetak{k}| \bbr)}\{(\thetak{k}{m,n}-\thetahatk{k}{m,n})(\thetak{k}{l,n}-\thetahatk{k}{l,n})\}.\nonumber\\
&& ~\mbox{for $n = 1$ and $m,l \in \{1, 2\}$}
\end{IEEEeqnarray}

Thus, the approximate MAP detector is written as
\begin{IEEEeqnarray}{rCl}\label{eq:estdetappr_a}
\hat{\bm c}_{k} &=& {\arg\underset{ \bck{k} \in \C}\max}  \underset{\bThetak{k}}{\int}  p(\brk{k}|\bck{k}, \bThetak{k}) p(\bThetak{k}| \bck{k}, \bbr_{k})  \tr d \bThetak{k},\\
\label{eq:estdetappr_b}
&\approx& {\arg\underset{ \bck{k} \in \C}\max}  \underset{\bThetak{k}}{\int}  p(\brk{k}|\bck{k}, \bThetak{k}) p(\bThetak{k}| \bbr)  \tr d \bThetak{k} \\
\label{eq:estdetappr}
&=& {\arg\underset{ \bck{k} \in \C}\max}  \underset{\bThetak{k}}\int p(\brk{k}|\bck{k},\bThetak{k}) \N(\bThetak{k};\bThetahatk{k},\mathbf{\hat{P}}_{k}) \tr d \bThetak{k}  \\
\label{eq:estdetappr_c}
&\approx&  {\arg\underset{ \bck{k} \in \C}\max} \; \exp\!\left\{-\frac{|\ck{k}{1}|^{2}+|\ck{k}{2}|^{2}}{N_{0}}\right\} \nonumber\\&& ~~~~~~~~~~\cdot \mbox{I}_{0}\left(|\uk{k}{1,1}| + |\uk{k}{2,1}| - |\utk{k}{1,2}| \right).
\end{IEEEeqnarray}
We refer the reader to Appendix B for the derivation of the result in \refeq{eq:estdetappr_c}. In \refeq{eq:estdetappr}, we define
\begin{IEEEeqnarray}{rCl}\label{eq:udefs}
\uk{k}{m,1} &\triangleq& \frac{2}{N_{0}}\rk{k}{1}{\ck{k}{m}}^{*} + \frac{  e^{\jmath \hat{\theta}_{k}^{(m,1)}}}{P_{k,1}^{(m,1)}}, ~\mbox{$m \in \{1, 2\}$} \nonumber\\
\utk{k}{1,2} &\triangleq& \frac{2}{N_{0}}\ck{k}{2}{\ck{k}{1}}^{*} +   \left| \uttk{k}{1,2}\right|e^{\jmath (\hat{\theta}_{k}^{(1,1)} - \hat{\theta}_{k}^{(2,1)})},
\end{IEEEeqnarray}
where $\left| \uttk{k}{1,2}\right|$ is obtained by solving \cite{mardia07}
\begin{IEEEeqnarray}{rCl}\label{eq:u3tildesolve}
P_{k,1}^{1,2} = \frac{-|\uttk{k}{1,2}|}{\sqrt{\left({P_{k,1}^{(1,1)}}^{-1} - |\uttk{k}{1,2}|\right)\left({P_{k,1}^{(2,2)}}^{-1} - |\uttk{k}{1,2}|\right)}}.
\end{IEEEeqnarray}

The generalization of the approximate MAP detector in \refeq{eq:estdetappr_c} to arbitrary $N_{\tr t}$ and $N_{\tr r}$ values is given as
\begin{IEEEeqnarray}{rCl}\label{eq:estdetappr_gen}
\hat{\bm c}_{k} &=& {\arg\underset{ \bck{k} } \max}  \; \exp\!{\left(-N_{\tr r}\underset{m=1}{\overset{N_{\tr t}}{\sum}}\frac{|\ck{k}{m}|^{2}}{N_{0}}\right)} \nonumber\\&& ~~~~~~~~~ \underset{n=1}{\overset{N_{\tr r}}{\prod}}  \mbox{I}_{0}\left(\underset{m=1}{\overset{N_{\tr t}}{\sum}} |\uk{k}{m,n}|  - \underset{\stackrel{m=1,}{l > m}}{\overset{N_{\tr t}}{\sum}}|\utk{k}{m,l}| \right).
\end{IEEEeqnarray}

For future reference, we refer to the smoother-detector algorithm for approximate MAP detection in \refeq{eq:estdetappr_c} based on the Gaussian pdf assumption for phase noise as Gauss-MAP. When the a posteriori pdf $p(\bThetak{k}| \bck{k}, \bbr_{k})$ is considered to be a Dirac Delta function, $\delta(\bThetak{k} - \bThetahatk{k})$, then the symbol detector in \refeq{eq:estdetappr} reduces to the Euclidean distance-based detector that treats the phase noise estimate  as the true value of the phase noise at the $k$th time instant, i.e.,
\begin{IEEEeqnarray}{rCl}\label{eq:estdeteuc}
\hat{\bm c}_{k} &=&  {\arg\underset{ \bck{k} \in \C }\max}  \underset{\bThetak{k}}\int p(\brk{k}|\bck{k},\bThetak{k}) \delta(\bThetak{k} - \bThetahatk{k}) \tr d \bThetak{k}  \nonumber\\
&=&  {\arg\underset{ \bck{k} \in \C }\max} \; p(\brk{k}|\bck{k},\bThetahatk{k}).
\end{IEEEeqnarray}
In the sequel, we refer to the symbol-by-symbol detector comprising the smoother and the Euclidean distance-based detector  in \refeq{eq:estdeteuc} as EUC-MAP.

\section{VB Framework-Based Algorithm for Approximate MAP Detection}
\label{sec:vb}
In this section, we develop a receiver algorithm for approximating the MAP symbol detector based on the VB framework as in \cite{nissila09} (see \cite{beal_thesis} for a nice tutorial on the VB framework). We consider arbitrary $N_{\tr t}$ and $N_{\tr r}$, and with a slight abuse of notation, we define $\bThetak{k} = [\thetak{k}{1,1}, \ldots,\thetak{k}{m,n},\ldots, \thetak{k}{N_{\tr t},N_{\tr r}}]$ and $\bck{k} = [\ck{k}{1}, \ldots, \ck{k}{N_{\tr t}}]$. Based on this framework, we first compute the log likelihood of $\bbr$ as
\begin{IEEEeqnarray}{rCl}\label{eq:loglikelihood_bbr}
\log p(\bbr) &=& \log \sum _{\bbc} \int_{\bbTheta} p(\bbc,\bbTheta,\bbr) \tr d \bbTheta\nonumber\\
&{=}& \log \sum _{\bbc} \int_{\bbTheta} Q(\bbc,\bbTheta)\frac{p(\bbc,\bbTheta,\bbr)}{Q(\bbc,\bbTheta)} \tr d \bbTheta \nonumber\\
&\overset{(\tr{a})}{\geq}&  \sum _{\bbc} \int_{\bbTheta} Q(\bbc,\bbTheta) \log \frac{p(\bbc,\bbTheta,\bbr)}{Q(\bbc,\bbTheta)}\tr d \bbTheta.
\end{IEEEeqnarray}
In \refeq{eq:loglikelihood_bbr}, the Jensen's inequality is applied to lower bound the log likelihood; when $Q(\bbc,\bbTheta)$ is set to $P(\bbc,\bbTheta|\bbr)$, the lower bound is achieved. Thus, our objective is to search over the various pdfs $Q(\bbc,\bbTheta)$ can assume, such that the bound in \refeq{eq:loglikelihood_bbr} is as tight as possible. In order to reduce the search space, as in \cite{beal_thesis, nissila09}, we constrain $Q(\bbc,\bbTheta)$ to a family of factorized pdfs, i.e., we assume that $Q(\bbc,\bbTheta) = q_{\mathbf{c}}(\bbc)q_{\boldsymbol{\theta}}(\bbTheta)$. This also corresponds to the assumption that $\bbc$ and $\bbTheta$ are independent of each other given $\bbr$. Hence, the lower bound is rewritten as
\begin{IEEEeqnarray}{rCl}\label{eq:loglikelihood_lb}
\log p(\bbr) &\geq&  \sum _{\bbc} \int_{\bbTheta} q_{\mathbf{c}}(\bbc)q_{\boldsymbol{\theta}}(\bbTheta) \log \frac{p(\bbc,\bbTheta,\bbr)}{q_{\mathbf{c}}(\bbc)q_{\boldsymbol{\theta}}(\bbTheta)} \tr d \bbTheta \nonumber\\
&\triangleq& \mathcal{H}(q_{\mathbf{c}}(\bbc),q_{\boldsymbol{\theta}}(\bbTheta),\bbr),
\end{IEEEeqnarray}
where $\mathcal{H}(q_{\mathbf{c}}(\bbc),q_{\boldsymbol{\theta}}(\bbTheta),\bbr)$ is referred to as the variational free energy -- its maximization results in the minimization of the KL divergence between $q_{\mathbf{c}}(\bbc)q_{\boldsymbol{\theta}}(\bbTheta)$ and $p(\bbc,\bbTheta|\bbr)$. To determine the factorized pdf, $q_{\mathbf{c}}(\bbc)$ and $q_{\boldsymbol{\theta}}(\bbTheta)$, that maximize $\mathcal{H}$, a coordinate ascent routine is used that maximizes with respect to one pdf while keeping the other fixed, in an alternating manner. Based on the functional derivatives of $\mathcal{H}$ with respect to the factorized pdf \cite{nissila09}, the coordinate ascent routine involves the iterative computation of
\begin{IEEEeqnarray}{rCl}\label{eq:pthetapc_update}
q_{\boldsymbol{\theta}}(\bbTheta) &\propto& p(\bbTheta)e^{\underset{\bbc}\sum q_{\mathbf{c}}(\bbc)\log P(\bbr|\bbc,\bbTheta)}  \nonumber\\ 
q_{\mathbf{c}}(\bbc) &\propto&  P(\bbc)e^{\int_{\boldsymbol{\bbTheta}} q_{\boldsymbol{\theta}}\bbTheta)\log P(\bbr|\bbc,\bbTheta)\tr d \bbTheta}.
\end{IEEEeqnarray}
The coordinate ascent routine is ensured to converge to a fixed point \cite{beal_thesis}, but global optimality is not guaranteed.

We can immediately see that the coordinate ascent routine results in a receiver algorithm that iteratively computes the a posteriori phase noise pdf and the symbol pmf as given in \refeq{eq:pthetapc_update}. To derive their respective functional forms, we consider the received signal model in \refeq{eq:received_mimo}. Based on \refeq{eq:pthetapc_update}, the factorized pdf of $\bbTheta$ is derived as
\begin{IEEEeqnarray}{rCl}\label{eq:q_theta}
q_{\boldsymbol{\theta}}(\bbTheta)  &\approx&  p(\bbTheta|\bbr,\underline{\mathbf{c}}), \textrm{ where, } \nonumber\\
\mathbb{E}_{q_{\mathbf{c}}} \{\bck{k}\} & = & \underline{\mathbf{c}}_{k},\, \underline{\mathbf{c}} = [\underline{\mathbf{c}}_{1}, \ldots,\underline{\mathbf{c}}_{L}], {\mathrm{Var}_{q_{\mathbf{c}}}}\{\bck{k}\}  \approx  0,
\end{IEEEeqnarray}
where $\underline{\mathbf{c}}$ denotes the sequence of symbol averages transmitted by all transmit antennas and is used for computing the factorized pdf $q_{\boldsymbol{\theta}}(\bbTheta)$. Furthermore, these symbol averages are treated as the true transmitted symbols as imposed by the variance constraint. We refer the reader to Appendix C for the proof of this result.

The factorized pmf of $\bbc$ is given by
\begin{IEEEeqnarray}{rCl}\label{eq:q_symbol}
q_{\mathbf{c}}(\bbc) &=& C_{\boldsymbol{c}}\prod_{k=1}^{L} P(\bck{k})\overset{N_{\tr r}}{\underset{n=1}{\prod}} \exp\!\{C_\tr{temp}^{(2)} \}\nonumber\\
C_\tr{temp}^{(1)}  & = & -\frac{1}{N_{0}} \left\{\overset{N_{\tr r}}{\underset{n=1}{\sum}} {\left|\rk{k}{n}-\overset{N_{\tr t}}{\underset{m=1}{\sum}}\ck{k}{m}e^{\jmath\thetahatk{k}{m,n}}\right|^{2}} -\overset{N_{\tr t}}{\underset{m=1}{\sum}}\left|\ck{k}{m}\right|^{2} \right.\nonumber\\
 && \left.\cdot P_{k,n}^{(m,m)}  -\overset{N_{\tr t}}{\underset{m=1}{\sum}}\underset{\underset{l\neq m}{l=1}}{\overset{N_{\tr t}}{\sum}}\ck{k}{m}{\ck{k}{l}}^{*}P_{k,n}^{(m,l)} e^{\jmath(\thetahatk{k}{m,n} - \jmath\thetahatk{k}{l,n})} \right\}. \IEEEeqnarraynumspace
\end{IEEEeqnarray}
In \refeq{eq:q_symbol}, we assumed that the a priori symbol sequence probability factorizes fully, which is reasonable in uncoded transmissions or in coded transmissions where pseudorandom interleavers are employed. The constant $C_{\boldsymbol{c}}$ normalizes the pmf and is independent of the transmitted symbols. The estimates of the phase noise in each link and the covariance matrix $\mathbf{P}_{k,n}$ are obtained by using an off-the-shelf smoother \cite{kay}.

Thus, the approximate MAP detector based on the VB framework is
\begin{IEEEeqnarray}{rCl} \label{eq:vb_map}
\hat{\bm c}_{k} &=& {\arg\underset{ \bck{k} }\max} \; q_{\mathbf{c}}(\bbc),
\end{IEEEeqnarray}
where $q_{\mathbf{c}}(\bbc)$ corresponds to the symbol pmf to which the coordinate ascent routine in \refeq{eq:pthetapc_update} converges. For future reference, we refer to the approximate MAP detector that is derived based on the VB framework as VB-MAP.

\section{Simulation Results}
\label{sec:sim_res}

\begin{figure}[!t]
\begin{center}
\includegraphics[width = 3.5in, keepaspectratio=true]{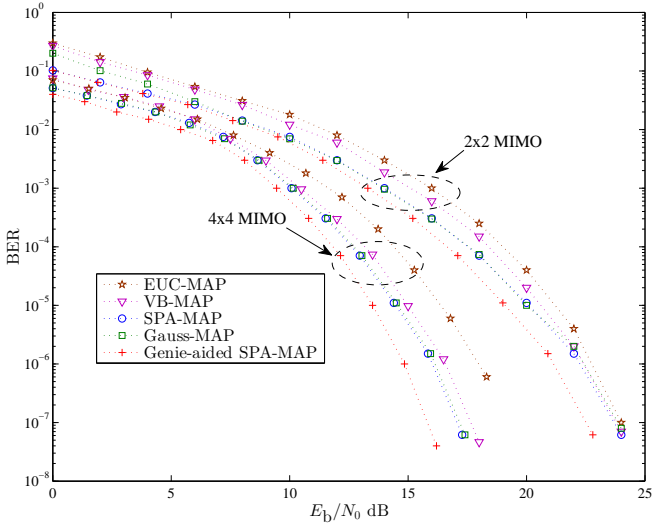}
\caption{BER performances for uncoded data transmission in different MIMO systems using the different receiver algorithms for $\sigmap{\tr{t}}{} = \sigmap{\tr{r}}{} = 4^{\circ}$.}
\label{fig:miso_diff_antennas}
\end{center}
\end{figure}

In this section we study the performances of the receiver algorithms proposed in the previous sections, namely the SPA-MAP \refeq{eq:puck_fstphn}, Gauss-MAP \refeq{eq:estdetappr_c} and VB-MAP \refeq{eq:q_symbol}, and those from prior work, namely the EUC-MAP \refeq{eq:estdeteuc}. The SPA-MAP algorithm with perfect symbol information in \refeq{eq:pdtheta} is considered as the benchmark algorithm and is referred to as the genie-aided SPA-MAP. That is, all the transmitted symbols are considered as pilots for phase estimation using \refeq{eq:afkupdate_fstphn}, \refeq{eq:abkupdate_fstphn} followed by symbol detection based on \refeq{eq:puck_fstphn}.

For implementing the Gauss-MAP, VB-MAP and EUC-MAP, the EKS is used to compute the estimate of the phase noise in each link and its respective variance. These estimates and their variances are used by the detector to compute the a posteriori symbol pmf, which we denote as $P_{\mathbf{c}}(\bck{k})$. For the Gauss-MAP, EUC-MAP and VB-MAP, $P_{\mathbf{c}}(\bck{k})$ is computed using \refeq{eq:estdetappr_c}, \refeq{eq:estdeteuc} and \refeq{eq:q_symbol}, respectively.  Then the symbol average and its variance are computed as  ${\mathbb{E}}_{ P_{\mathbf{c}}} \{\bck{k}\}  =  \underline{\mathbf{c}}_{k},$  $\mathrm{Var}_{P_{\mathbf{c}}}\{\bck{k}\}=\sigmap{\tr c}{2}$, and these symbol statistics are conveyed back to the EKS. The linearized state space model of the EKS is derived by modeling the symbol transmitted by the $m$th transmit antenna at the $k$th time instant as
\begin{IEEEeqnarray}{rCl} \label{eq:soft_sym}
\ck{k}{m} = \underline{c}_{k}^{(m)} + \wk{\tr c,k}{m}.
\end{IEEEeqnarray}
In \refeq{eq:soft_sym}, $\underline{c}_{k}^{(m)}$ is the symbol average from the detector and $\wk{\tr c,k}{m}$ is the error associated with it, which is assumed to be Gaussian distributed, i.e., $\wk{\tr c,k}{m} \sim \N(0,\sigmap{\tr c}{2})$ \cite{reza13, singer07}. Thus the state space model considering the received signal at the $n$th receive antenna and the $k$th time instant is given as
\begin{eqnarray}\label{eq:ekf_ssm}
\rk{k}{n} &=&  \overset{N_{\tr t}}{\underset{m=1}{\sum}} e^{\jmath {\thetak{k}{m,n}}}(\underline{c}_{k}^{(m)} + \wk{\tr c,k}{m}) +  \wk{k}{n}\nonumber\\
&\approx&  \overset{N_{\tr t}}{\underset{m=1}{\sum}} e^{\jmath {\thetahatk{k}{m,n}}}(1+\jmath (\thetak{k}{m,n}-\thetahatk{k}{m,n}))\underline{c}_{k}^{(m)} + \tilde{w}_{k}^{(n)}\nonumber\\
 \thetak{k}{m,n} &=& \thetak{k-1}{m,n} + \Deltakn{\tr{t},k}{m} + \Deltakn{\tr{r},k}{m},
\end{eqnarray}
where $\tilde{w}_{k}^{(n)} \sim \N(0,N_{0} + \sigmap{\tr c}{2})$.

In uncoded transmission, we perform $2$ iterations between the smoother and the detector, beyond which the performance gain is observed to be marginal. After reaching the maximum number of iterations, the transmitted symbols are decided as $\hat{\bm c}_{k} = {\argmax}_{ \bck{k} } \; P_{\mathbf{c}}(\bck{k})$. For evaluating the performance of the algorithms in this transmission mode, we consider binary phase-shift keying (BPSK) unless otherwise stated and the length of a data frame is $L = 10000$ symbols. Furthermore, we place $10$ consecutive pilot symbols at the beginning of each frame, and $1$ pilot symbol every $20$ data symbols yielding a pilot density of around $5.1\%$.

In coded transmission, the symbol pmf computed by the detector is used by the decoder for computing the bit LLRs, and $2$ global iterations are performed between the detector and the decoder, beyond which the performance gain is seen to be marginal. After the maximum number of iterations is reached, $P_{\mathbf{c}}(\bck{k})$ is used by the decoder to make hard decisions on the information bits. Note that in this transmission mode, using EUC-MAP as the detector corresponds to the turbo-synchronization algorithm for MIMO systems proposed in \cite{ali13, guido13}. We consider rate-$1/2$ and rate-$4/5$  low-density parity-check (LDPC) codes of length $L=64800$. The pilot distribution and the modulation scheme employed are the same as the uncoded transmission case unless otherwise stated.

\begin{figure}[!t]
\begin{center}
\includegraphics[width = 3.5in, keepaspectratio=true]{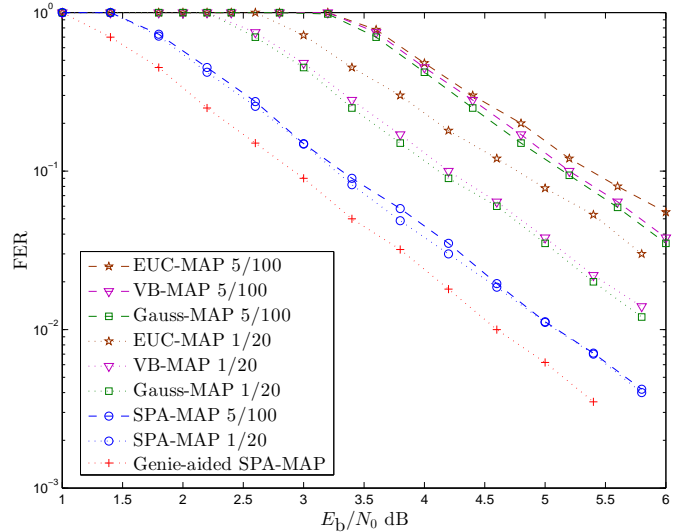}
\caption{FER performances of a $2 \times 1$ system using different receiver algorithms for code rate $= 1/2$, $\sigmap{\tr{t}}{} = \sigmap{\tr{r}}{} = 4^{\circ}$ and different pilot distributions.}
\label{fig:diff_rates}
\end{center}
\end{figure}

We consider data transmission (both uncoded and coded) in a strong phase noise scenario that corresponds to $\sigmap{\tr{t}}{} = \sigmap{\tr{r}}{} = 4^{\circ}$ \cite{colavolpe05}.  The channel is considered to be Rayleigh fading and is assumed to be known (estimated). Different channel realizations are generated for each data frame. First, we investigate the performances of the proposed algorithms for different values of $N_{\tr t}$ and $N_{\tr r}$. In Fig.~\ref{fig:miso_diff_antennas}, the bit error rate (BER) performance of the proposed algorithms is illustrated for $2 \times 2$ and $4 \times 4$ systems for uncoded transmission. For the $2 \times 2$ system, we note that all the proposed algorithms outperform the EUC-MAP for low-to-medium values of SNR per bit ($E_{\tr b}/N_{0}$). We observe that the SPA-MAP performs better than the Gauss-MAP especially for low values of $E_{\tr b}/N_{0}$ by around $1.5$ dB. Both Gauss-MAP and SPA-MAP perform better than the VB-MAP for low-to-medium values of $E_{\tr b}/N_{0}$ by around $1$ dB. For high $E_{\tr b}/N_{0}$ values, it can be seen that all algorithms perform similarly. Furthermore, we observe that the gap in the performance between the benchmark algorithm, the proposed algorithms and the EUC-MAP increases as both $N_{\tr t}$ and $N_{\tr r}$ are increased to $4$. In particular, the gap between the proposed algorithms and the EUC-SPA is around $2$ dB for high values of $E_{\tr b}/N_{0}$. This can be attributed to higher amplitude distortions due to phase noise experienced  by the transmitted symbols as the number of antennas increases \cite{rajet12}.

Next, in Fig.~\ref{fig:diff_rates}, we investigate the frame error rate (FER) performance of the proposed algorithms for the coded transmission mode considering rate-$1/2$ LDPC code and  a $2 \times 1$ system. We maintain an overall pilot density of $5.1\%$, and we consider two pilot-symbol distributions -- $1/20$ denotes the arrangement where a pilot symbol is placed every $20$ symbols, and $5/100$ indicates that $5$ consecutive pilot symbols are placed every $100$ symbols. Specifically, for both pilot distributions, we observe that the SPA-MAP performs better than all the other algorithms. For the $1/20$ pilot distribution case, we observe that the SPA-MAP outperforms all the other algorithms by $0.7$ dB, and for the $5/100$ pilot distribution case, the gap in performance widens to $1.5$ dB. We note that the Gauss-MAP, VB-MAP and EUC-MAP are more prone to estimation errors when the SNR and hence the energy of the pilot symbols is low. Furthermore, the EKS-based receiver algorithms are seen to be extremely sensitive to the pilot-symbol arrangement, while the SPA-MAP is seen to be the least sensitive.

\begin{figure}[!t]
\begin{center}
\includegraphics[width = 3.5in, keepaspectratio=true]{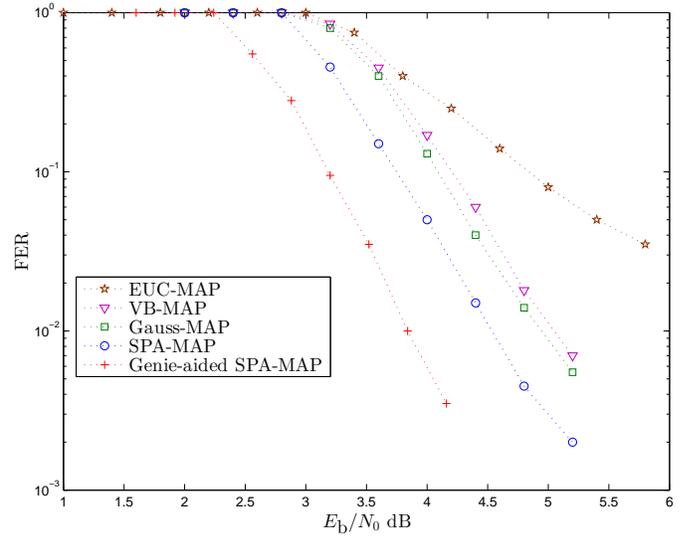}
\caption{FER performance of a $2 \times 1$ system using different receiver algorithms for code rate $= 4/5$ and $\sigmap{\tr{t}}{} = \sigmap{\tr{r}}{} = 4^{\circ}$.}
\label{fig:diff_rates_4by5}
\end{center}
\end{figure}

In Fig.~\ref{fig:diff_rates_4by5}, we evaluate the FER performance of all algorithms for rate-$4/5$ LDPC code considering a $2 \times 1$ system in order to study the dependence of the performance of the algorithms on the code rate. We observe that the SPA-MAP outperforms all other algorithms by a significant margin and the gap in the performance between SPA-MAP and the EKS-based detectors decreases with increasing code rate (as compared to Fig.~\refeq{fig:diff_rates}). This is because the pilot symbols have higher energy when the code rate is higher and are thus more reliable.

\begin{figure}[!t]
\begin{center}
\includegraphics[width = 3.5in, keepaspectratio=true]{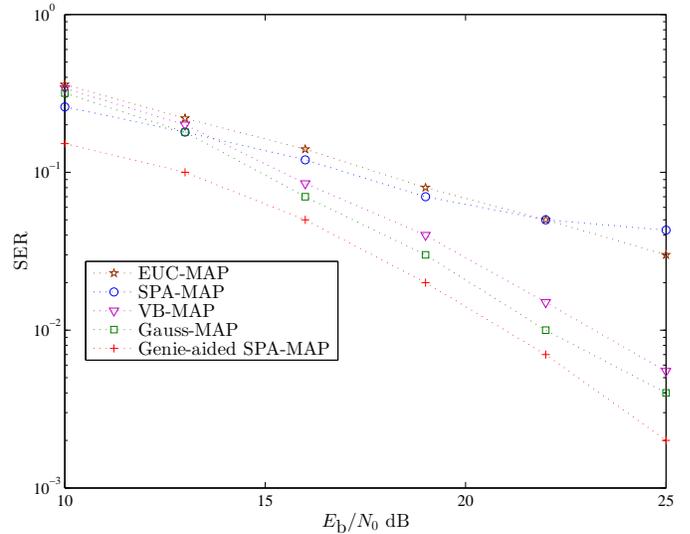}
\caption{SEP performance of a $2 \times 1$ system  for $16$-QAM transmission  using different receiver algorithms for $\sigmap{\tr{t}}{} = \sigmap{\tr{r}}{} = 4^{\circ}$.}
\label{fig:diff_const}
\end{center}
\end{figure}

\begin{figure}[!t]
\begin{center}
\includegraphics[width = 3.5in, keepaspectratio=true]{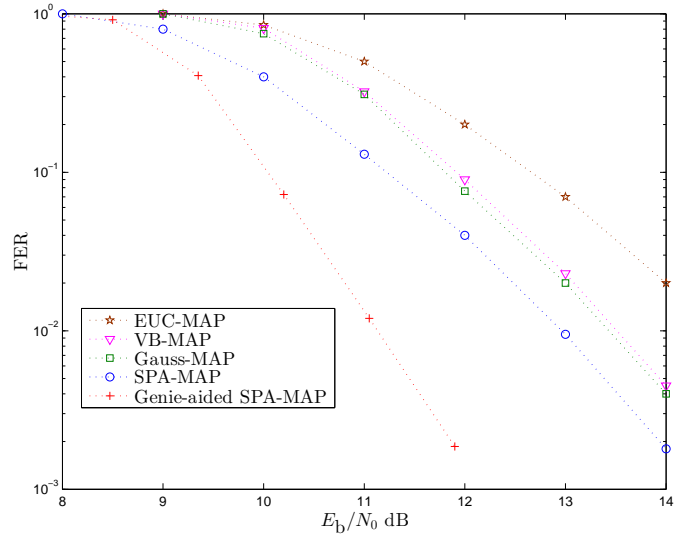}
\caption{FER performance of a $2 \times 1$ system  for $16$-QAM transmission, for code rate $= 4/5$  using different receiver algorithms for $\sigmap{\tr{t}}{} = \sigmap{\tr{r}}{} = 4^{\circ}$.}
\label{fig:diff_const_coded}
\end{center}
\end{figure}

Finally, we analyze the performance of the algorithms for different constellation sizes. The symbol error rate (SER) performance of the algorithms is presented in Fig. \ref{fig:diff_const} for uncoded $16$-QAM transmission. We observe that the Gauss-MAP and the VB-MAP outperform the SPA-MAP. This owes to the approximation of the Tikhonov mixture to a single mode Tikhonov pdf in \refeq{eq:pdtheta} using uniform a priori information for the transmitted symbols in $\Pdc(\bck{k})$. This approximation can be highly erroneous for large non-equal energy constellations particularly when reliable a priori information of the transmitted symbols is not available to the detector. However, for the case of coded transmission considering rate-$4/5$ LDPC code in Fig. \ref{fig:diff_const_coded}, the FER performance of the SPA-MAP algorithm is seen to be superior to all the other algorithms. This is because the LDPC decoder provides a more reliable a priori information in $\Pdc(\bck{k})$ of the transmitted symbols rendering the single mode Tikhonov pdf approximation more accurate.

\section{Conclusions}
\label{sec:concl}

In this paper, we derived the optimum MAP symbol detector that involves the joint estimation of the a posteriori phase noise pdf and data detection. The optimum receiver structure is seen to be intractable and unimplementable in practice, since the exact phase noise pdf computation is an infinite dimensional problem. In this regard, we proposed three suboptimal, low-complexity algorithms that were observed to outperform all the existing techniques in the literature. In particular, the receiver based on the sum-product algorithm, SPA-MAP, was found to perform better than all the other algorithms for both uncoded and coded transmission of BPSK symbols. For higher-order constellations (16-QAM), the algorithms based on the smoother-detector structure, Gauss-MAP, and the variational Bayesian framework, VB-MAP, were observed to perform the best in the case of uncoded transmission. However, for coded $16$-QAM tranmission, the SPA-MAP algorithm was seen to be superior to all the other algorithms considered. Finally, we observed that the SPA-MAP is less sensitive to pilot symbol placements as opposed to the algorithms that use an EKS for computing the a posteriori phase noise pdf.

\section*{APPENDIX A}  
\section*{Derivation of the SPA Messages and Computation of their Parameters}
\label{sec:app_a}

The message $\Pftheta(\bThetak{k})$ is derived for the case of the Wiener phase noise process by evaluating \refeq{eq:pfthetak} using the approximation \cite[eq. (42)]{colavolpe05}
\begin{IEEEeqnarray}{rCl} \label{eq:cola_appr}
\frac{1}{\sqrt{2\pi \sigma^{2}}}\int_{0}^{2\pi} && e^{\Re\left[ze^{-\jmath  \varphi }\right]} e^{\frac{-(\varphi - \phi)^{2}}{2\sigma^{2}}}  \tr d  \varphi \stackrel{\sim}{\propto} \exp\!\left\{ \Re\left[\frac{ze^{-\jmath   \phi}}{1+|z|\sigma^{2}}\right]\right\} \nonumber\\
\end{IEEEeqnarray}
for $z \in \mathbb{C},$ $\sigma^{2}\in \mathbb{R}^{+},$ and $\varphi, \phi \in \mathbb{R}$. We first evaluate
\begin{IEEEeqnarray}{rCl}\label{eq:a1temp}
&& A_{\tr{temp}}^{(1)}\nonumber\\
&& = \int_{0}^{2\pi} \exp\!\left\{ \Re\left[(\yk{k-1}{1}e^{-\jmath \thetak{\tr{t},k-1}{1}}+\yk{k-1}{2}e^{-\jmath\thetak{\tr{t}, k-1}{2}})e^{-\jmath\thetak{\tr{r}, k-1}{1}} \right.\right.\nonumber\\&&\left.\left. ~~~~~~~~~~~~~~-\yk{k-1}{3}e^{-\jmath(\thetak{\tr{t}, k-1}{1} - \thetak{\tr{t}, k-1}{2})}\right]\right\}\nonumber\\ && ~~~~~~~~~\cdot \Pdelta(\thetak{\tr{r}, k}{1}-\thetak{\tr{r}, k-1}{1}) \tr d \thetak{\tr{r}, k-1}{1}  \IEEEeqnarraynumspace\\
&& =  \exp\!\left\{ -\Re\left[\yk{k-1}{3}e^{-\jmath(\thetak{\tr{t},k-1}{1}-\thetak{\tr{t},k-1}{2})}\right]\right\} \nonumber\\&& ~~~\cdot\int_{0}^{2\pi} \exp\!\left\{ \Re\left[\left(\yk{k-1}{1}e^{-\jmath \thetak{\tr{t},k-1}{1}}+\yk{k-1}{2}e^{-\jmath\thetak{\tr{t},k-1}{2}}\right) e^{-\jmath \thetak{\tr{r},k-1}{1}} \right]\right\}\nonumber\\ &&  ~~~~~\cdot\Pdelta(\thetak{\tr{r}, k}{1} - \thetak{\tr{r}, k-1}{1}) \tr d \thetak{\tr{r}, k-1}{1}  \nonumber\\
 \label{eq:cola_appr_a}
&&\propto \exp\!\left\{ -\Re\left[\yk{k-1}{3}e^{-\jmath(\thetak{\tr{t},k-1}{1}-\thetak{\tr{t},k-1}{2})}\right]\right\}  \nonumber\\&&  ~~~\cdot\exp\!\left\{ \Re\left[\frac{\left(\yk{k-1}{1}e^{-\jmath \thetak{\tr{t},k-1}{1}}+\yk{k-1}{2}e^{-\jmath\thetak{\tr{t},k-1}{2}}\right)e^{-\jmath\thetak{\tr{r},k}{1}}}{1+\sigmap{\tr{r}}{2}\left|\yk{k-1}{1}e^{-\jmath\thetak{\tr{t},k-1}{1}}+\yk{k-1}{2}e^{-\jmath\thetak{\tr{t},k-1}{2}}\right|}\right]\right\} \\
\label{eq:cola_appr_b}
&&\approx \exp\!\left\{ -\Re\left[\yk{k-1}{3}e^{-\jmath(\thetak{\tr{t},k-1}{1}-\thetak{\tr{t},k-1}{2})}\right]\right\} \nonumber\\&&  \cdot \exp\!\left\{ \Re\left[\frac{\left(\yk{k-1}{1}e^{-\jmath\thetak{\tr{t},k-1}{1}}+\yk{k-1}{2}e^{-\jmath\thetak{\tr{t},k-1}{2}}\right)}{1+\sigmap{\tr{r}}{2}\left(|\yk{k-1}{1}|+|\yk{k-1}{2}|\right)}e^{-\jmath\thetak{\tr{r},k}{1}}\right]\right\}.
\end{IEEEeqnarray}
In \refeq{eq:cola_appr_a}, we used the approximation in \refeq{eq:cola_appr}, and to obtain the result in \refeq{eq:cola_appr_b} we applied the approximation
\begin{IEEEeqnarray}{rCl}\label{eq:ya1_appr}
&& \left|\yk{k-1}{1}e^{-\jmath\thetak{\tr{t},k-1}{1}}+\yk{k-1}{2}e^{-\jmath\thetak{\tr{t},k-1}{2}}\right| \nonumber\\  && =  \left|\left|\yk{k-1}{1}\right|e^{\jmath(\angle \yk{k-1}{1} - \thetak{\tr{t},k-1}{1} - \thetak{\tr{r},k-1}{1})} + \left|\yk{k-1}{2}\right|e^{\jmath(\angle \yk{k-1}{2} - \thetak{\tr{r},k-1}{1} -  \thetak{\tr{t},k-1}{2})}\right| \nonumber\\ &&\approx \left|\yk{k-1}{1}\right|+\left|\yk{k-1}{2}\right|,
\end{IEEEeqnarray}
where it is assumed that $\angle \yk{k-1}{2} - \thetak{\tr{r},k-1}{1} -  \thetak{\tr{t},k-1}{2}$ and $\angle \yk{k-1}{1} - \thetak{\tr{r},k-1}{1} -  \thetak{\tr{t},k-1}{1}$ are very small, i.e., we assume that the difference between the phase noise in each link and its (predicted and updated) estimate is small. Now, define
\begin{IEEEeqnarray}{rCl}\label{eq:ydef}
\ytk{k-1}{1} \triangleq \frac{\yk{k-1}{1}e^{-\jmath\thetak{\tr{r},k}{1}}}{1+\sigmap{\tr{r}}{2}\left(|\yk{k-1}{1}|+|\yk{k-1}{2}|\right)}\nonumber\\
\ytk{k-1}{2} \triangleq \frac{\yk{k-1}{2}e^{-\jmath\thetak{\tr{r},k}{1}}}{1+\sigmap{\tr{r}}{2}\left(|\yk{k-1}{1}|+|\yk{k-1}{2}|\right)}.
\end{IEEEeqnarray}
Then compute
\begin{IEEEeqnarray}{rCl}\label{eq:a2temp}
A_{\tr{temp}}^{(2)} &\triangleq& \int_{0}^{2\pi} A_{\tr{temp}}^{(1)} \, \Pdelta(\thetak{\tr{t},k}{1} -\thetak{\tr{t},k-1}{1})\tr d\thetak{\tr{t},k-1}{1} \nonumber\\
&=&\int_{0}^{2\pi}    \exp\! \left\{ \Re \left[\ytk{k-1}{1}e^{-\jmath\thetak{\tr{t},k-1}{1}} + \ytk{k-1}{2}e^{-\jmath\thetak{\tr{t},k-1}{2}} \right.\right.\nonumber\\
&&\left.\left. ~~~~~~~~~~~ - \yk{k-1}{3}e^{-\jmath(\thetak{\tr{t},k-1}{1}-\thetak{\tr{t},k-1}{2})}\right]\right\} \nonumber\\ && ~~~~~~\cdot \Pdelta(\thetak{\tr{t},k}{1}- \thetak{\tr{t},k-1}{1})\tr d\thetak{\tr{t},k-1}{1} \nonumber\\
&=&\exp\!\left\{ \Re\left[\ytk{k-1}{2}e^{-\jmath\thetak{\tr{t},k-1}{2}}\right]\right\} \nonumber\\
&&\cdot \int\exp\!\left\{ \Re\left[\left(\ytk{k-1}{1}-\yk{k-1}{3}e^{\jmath\thetak{\tr{t},k-1}{2}}\right)e^{-\jmath\thetak{\tr{t},k-1}{1}}\right]\right\} \nonumber\\ &&~~~\cdot\Pdelta(\thetak{\tr{t},k}{1}- \thetak{\tr{t},k-1}{1})\tr d\thetak{\tr{t},k-1}{1}\nonumber\\
\label{eq:a2temp_a}
&\propto& \exp\!\left\{ \Re\left[\ytk{k-1}{2}e^{-\jmath\thetak{\tr{t},k-1}{2}}\right]\right\} \nonumber\\
&&\cdot \exp\!\left\{ \Re\left[\frac{\left(\ytk{k-1}{1}-\yk{k-1}{3}e^{\jmath\thetak{\tr{t},k-1}{2}}\right)e^{-\jmath\thetak{\tr{t},k}{1}}}{1+\sigmap{\tr{t}}{2}\left|\ytk{k-1}{1}-y_{3}e^{\jmath\thetak{\tr{t},k-1}{2}}\right|}\right]\right\}\\
\label{eq:a2temp_b}
&\approx& \exp\!\left\{ \Re\left[\ytk{k-1}{2}e^{-\jmath\thetak{\tr{t},k-1}{2}}\right]\right\} \nonumber\\
&&\cdot \exp\!\left\{ \Re\left[\frac{\left(\ytk{k-1}{1}-\yk{k-1}{3}e^{\jmath\thetak{\tr{t},k-1}{2}}\right)e^{-\jmath\thetak{\tr{t},k}{1}}}{1+\sigmap{\tr{t}}{2}\left|\left|\ytk{k-1}{1}\right|-\left|\yk{k-1}{3}\right|\right|}\right]\right\},
\end{IEEEeqnarray}
where in \refeq{eq:a2temp_a}, the approximation from \refeq{eq:cola_appr} is used. In \refeq{eq:a2temp_b}, we apply
\begin{IEEEeqnarray}{rCl}\label{eq:ya2_apprx}
&& \left|\ytk{k-1}{1}-\yk{k-1}{3}e^{\jmath\thetak{\tr{t},k-1}{2}}\right| \nonumber\\
&& = \left|\left|\ytk{k-1}{1}\right|e^{\jmath\angle \ytk{k-1}{1}}-\left|\yk{k-1}{3}\right|e^{\jmath\angle \yk{k-1}{3}}e^{\jmath\thetak{\tr{t},k-1}{2}}\right|\nonumber\\
&& \simeq\left|\left|\ytk{k-1}{1}\right|e^{\jmath\angle \ytk{k-1}{1}}-\left|\yk{k-1}{3}\right|e^{\jmath(\angle \yk{k-1}{1} - \angle \yk{k-1}{2} + \thetak{\tr{t},k-1}{2})}\right|\nonumber\\
&&  = \left|\left|\ytk{k-1}{1}\right|e^{\jmath\angle \ytk{k-1}{1}}-\left|\yk{k-1}{3}\right|e^{\jmath(\angle \ytk{k-1}{1} - \angle \yk{k-1}{2} + \thetak{\tr{r},k}{1} + \thetak{\tr{t},k-1}{2})}\right|\nonumber\\
&& \approx \left|\left|\ytk{k-1}{1}\right|e^{\jmath\angle \ytk{k-1}{1}}-\left|\yk{k-1}{3}\right|e^{\jmath \angle \ytk{k-1}{1}}\right| = \left|\left|\ytk{k-1}{1}\right|-\left|\yk{k-1}{3}\right|\right|, \nonumber\\
\end{IEEEeqnarray}
where in \refeq{eq:ya2_apprx}, it is considered that $\thetak{\tr{r},k}{1} + \thetak{\tr{t},k-1}{2} - \angle \yk{k-1}{2}$ is very small. Finally, we compute
\begin{IEEEeqnarray}{rCl}\label{eq:a3temp}
A_{\tr{temp}}^{(3)} &\triangleq& \int_{0}^{2\pi} A_{\tr{temp}}^{(2)} \, \Pdelta(\thetak{\tr{t},k}{2} - \thetak{\tr{t},k-1}{2})\tr d\thetak{\tr{t},k-1}{2} \nonumber\\
&=& \int_{0}^{2\pi} \exp\!\left\{ \Re\left[\ytk{k-1}{2}e^{-\jmath\thetak{\tr{t},k-1}{2}}\right]\right\} \nonumber\\ && ~~~~~\cdot \exp\!\left\{ \Re\left[\frac{\left(\ytk{k-1}{1}-\yk{k-1}{3}e^{\jmath\thetak{\tr{t},k-1}{2}}\right)}{1+\sigmap{\tr{t}}{2}\left|\left|\ytk{k-1}{1}\right|-\left|\yk{k-1}{3}\right|\right|}e^{-\jmath\thetak{\tr{t},k}{1}}\right]\right\}\nonumber\\
&& ~~~~~~\cdot \Pdelta(\thetak{\tr{t},k}{2} - \thetak{\tr{t},k-1}{2})\tr d \thetak{\tr{t},k-1}{2}\nonumber\\
&=&\exp\!\left\{ \Re\left[\frac{\ytk{k-1}{1}e^{-\jmath \thetak{\tr{t},k}{1}}}{1+\sigmap{\tr{t}}{2}\left|\left|\ytk{k-1}{1}\right|-\left|\yk{k-1}{3}\right|\right|}\right]\right\} \nonumber\\ &&\cdot \int_{0}^{2\pi}\exp\!\left\{ \Re\left[\tilde{\tilde{y}}_{k-1}^{(2)}e^{-\jmath\thetak{\tr{t},k-1}{2}}\right]\right\}\Pdelta(\thetak{\tr{t},k}{2} - \thetak{\tr{t},k-1}{2})\tr d \thetak{\tr{t},k-1}{2}\nonumber\\
&\propto& \exp\!\left\{ \Re\left[\frac{\ytk{k-1}{1}e^{-\jmath \thetak{\tr{t},k}{1}}}{1+\sigmap{\tr{t}}{2}\left|\left|\ytk{k-1}{1}\right|-\left|\yk{k-1}{3}\right|\right|} + \frac{\tilde{\tilde{y}}_{k-1}^{(2)}e^{-\jmath\thetak{\tr{t},k}{2}}}{1+\sigmap{\tr{t}}{2}\left|\tilde{\tilde{y}}_{k-1}^{(2)}\right|}\right]\right\} \nonumber \\
&\approx& \exp\!\left\{ \Re\left[\frac{\ytk{k-1}{1}e^{-\jmath \thetak{\tr{t},k}{1}}}{1+\sigmap{\tr{t}}{2}\left|\left|\ytk{k-1}{1}\right|-\left|\yk{k-1}{3}\right|\right|} \right.\right. \nonumber\\ &&~~~~~~ \left.\left. + \frac{\tilde{\tilde{y}}_{k-1}^{(2)}e^{-\jmath\thetak{\tr{t},k}{2}}}{1+\sigmap{\tr{t}}{2}\left|\left|\ytk{k-1}{2}\right|- \left|\yk{k-1}{3}\right|\right|}\right]\right\}, \; \mbox{where}\\
\tilde{\tilde{y}}_{k-1}^{(2)} &\triangleq& \ytk{k-1}{2}-\frac{{\yk{k-1}{3}}^{*}e^{\jmath\thetak{\tr{t},k}{1}}}{1+\sigmap{\tr{t}}{2}\left|\left|\ytk{k-1}{1}\right|-\left|\yk{k-1}{3}\right|\right|}. \nonumber
\end{IEEEeqnarray}
To obtain \refeq{eq:a3temp}, we apply the approximation from \refeq{eq:cola_appr}, and further we apply an approximation similar to that used in \refeq{eq:ya2_apprx}. Observe that the message $\Pftheta(\bThetak{k}) = A_{\tr{temp}}^{(3)}$, and hence
\begin{IEEEeqnarray}{rCl}\label{eq:pfthetafinal}
&& \Pftheta(\bThetak{k}) \nonumber\\
&&\approx  \exp\!\left\{ \Re\left[   \frac{\ytk{k-1}{1}e^{-\jmath \thetak{\tr{t},k}{1}}}{1+\sigmap{\tr{t}}{2}\left|\left|\ytk{k-1}{1}\right|-\left|\yk{k-1}{3}\right|\right|} + \frac{\ytk{k-1}{2}e^{-\jmath\thetak{\tr{t},k}{2}}}{1+\sigmap{\tr{t}}{2}\left|\left|\ytk{k-1}{2}\right| - \left|\yk{k-1}{3}\right|\right|}  \right.\right. \nonumber\\
&&\left.\left. -
\frac{{\yk{k-1}{3}}e^{-\jmath\thetak{\tr{t},k}{1}}}{\left(1+\sigmap{\tr{t}}{2}\left|\left|\ytk{k-1}{2}\right|- \left|\yk{k-1}{3}\right|\right|\right)\left(1+\sigmap{\tr{t}}{2}\left|\left|\ytk{k-1}{1}\right|-\left|\yk{k-1}{3}\right|\right|\right)}\right]\right\} \nonumber\\
&&\triangleq
\exp\!\left\{ \Re\left[(\afk{k}{1,1}e^{-\jmath \thetak{\tr{t},k}{1}}+\afk{k}{2,1}e^{-\jmath\thetak{\tr{t}, k}{2}})e^{-\jmath\thetak{\tr{r}, k}{1}} -\tafk{k}{1,2} \right.\right. \nonumber\\
&& \left.\left. ~~~~~~~~~~~\cdot e^{-\jmath(\thetak{\tr{t}, k}{1} - \thetak{\tr{t}, k}{2})}\right]\right\}.
\end{IEEEeqnarray}
From the result in \refeq{eq:pfthetafinal}, we arrive at the forward recursions presented in \refeq{eq:afkupdate_fstphn}. Computation of the message $\Pbtheta(\bThetak{k})$ to determine the backward parameter update equations in \refeq{eq:abkupdate_fstphn} proceeds similarly.

For computing the message $\Puc(\bck{k})$ in \refeq{eq:puck_fstphn}, define
\begin{IEEEeqnarray}{rCl}\label{eq:a4temp}
A_{\tr{temp}}^{(4)} &\triangleq& \int \exp\!\left\{ \Re\left[(\zk{k}{1,1}e^{-\jmath \thetak{\tr{t},k}{1}}+\zk{k}{2,1}e^{-\jmath\thetak{\tr{t}, k}{2}})e^{-\jmath\thetak{\tr{r}, k}{1}}   \right.\right. \nonumber\\ &&\left.\left. ~~~~~~~~-\tzk{k}{1,2}e^{-\jmath(\thetak{\tr{t}, k}{1} - \thetak{\tr{t}, k}{2})}\right]\right\} \tr d \bThetak{k}.
\end{IEEEeqnarray}
The integral in \refeq{eq:a4temp} is evaluated as
\begin{IEEEeqnarray}{rCl}
A_{\tr{temp}}^{(4)} & = & \int_{0}^{2\pi}  \int_{0}^{2\pi} \int_{0}^{2\pi}\exp\!\left\{ \Re\left[(\zk{k}{1,1}e^{-\jmath \thetak{\tr{t},k}{1}}+\zk{k}{2,1}e^{-\jmath\thetak{\tr{t}, k}{2}}) \right.\right. \nonumber\\ && \left.
\left.  \cdot  e^{-\jmath\thetak{\tr{r}, k}{1}}   -\tzk{k}{1,2}e^{-\jmath(\thetak{\tr{t}, k}{1} - \thetak{\tr{t}, k}{2})}\right]\right\}  \tr d \thetak{\tr{r}, k}{1}  \tr d \thetak{\tr{t}, k}{2}\tr d \thetak{\tr{t}, k}{1} \nonumber\\
 &=&    \int_{0}^{2\pi} \int_{0}^{2\pi} \exp\!\left\{ -\Re\left[\tzk{k}{1,2}e^{-\jmath(\thetak{\tr{t},k}{1}-\thetak{\tr{t},k}{2})}\right]\right\}  \nonumber\\ && \cdot \int_{0}^{2\pi}\exp\!\left\{ \Re\left[\left(\zk{k}{1,1}e^{-\jmath \thetak{\tr{t},k}{1}}+\zk{k}{2,1}e^{-\jmath\thetak{\tr{t},k}{2}}\right)e^{-\jmath \thetak{\tr{r},k}{1}}\right]\right\} \nonumber\\ && ~~~~~ \tr d \thetak{\tr{r}, k}{1} \tr d \thetak{\tr{t}, k}{2}\tr d \thetak{\tr{t}, k}{1}   \nonumber\\
\label{eq:a4tempfinal_a}
 &\propto&   \int_{0}^{2\pi}  \int_{0}^{2\pi} \exp\!\left\{ -\Re\left[\tzk{k}{1,2}e^{-\jmath(\thetak{\tr{t},k}{1}-\thetak{\tr{t},k}{2})}\right]\right\} \nonumber\\ &&  \cdot  \tr I_{0} \left( \left|\zk{k}{1,1}e^{-\jmath \thetak{\tr{t},k}{1}}+\zk{k}{2,1}e^{-\jmath\thetak{\tr{t},k}{2}}\right|\right) \tr d \thetak{\tr{t}, k}{2}\tr d \thetak{\tr{t}, k}{1} \\
 \label{eq:a4tempfinal_b}
 &\approx&    \int_{0}^{2\pi}  \int_{0}^{2\pi} \exp\!\left\{ -\Re\left[\tzk{k}{1,2}e^{-\jmath(\thetak{\tr{t},k}{1}-\thetak{\tr{t},k}{2})}\right]\right\} \nonumber\\&& \cdot  \tr I_{0} \left( \left|\left|\zk{k}{1,1}\right| +\left|\zk{k}{2,1}\right|\right|\right)  \tr d \thetak{\tr{t}, k}{2}\tr d \thetak{\tr{t}, k}{1} \\
 \label{eq:a4tempfinal_c}
  &\propto&  \tr I_{0}\left(\left|\tzk{k}{1,2}\right|\right) \tr I_{0} \left( \left|\left|\zk{k}{1,1}\right| +\left|\zk{k}{2,1}\right|\right|\right),
\end{IEEEeqnarray}
where in \refeq{eq:a4tempfinal_a} the result follows from \refeq{eq:cola_appr}, and the result in \refeq{eq:a4tempfinal_b} is obtained by applying the approximation in \refeq{eq:ya1_appr}. Further, applying the Tikhonov normalization constant, the result in \refeq{eq:a4tempfinal_c} is obtained. The result in \refeq{eq:a4tempfinal_c} is used in \refeq{eq:a4temp} to yield $\Puc(\bck{k})$  in  \refeq{eq:puck_fstphn}.

The generalizations presented in \refeq{eq:afkupdate_gen} and \refeq{eq:abkupdate_gen} are obtained by first identifying that $\Pdtheta(\bThetak{k})$ fully factorizes in terms of the receive antenna index. Further, the Gaussian pdf that minimizes the KL divergence with respect to $\Pdtheta(\bThetak{k})$ is obtained by performing moment matching with each of its factors \cite{minka05}. Following this, the remaining steps proceed similarly as presented in \refeq{eq:a1temp}-\refeq{eq:a4tempfinal_c}.

\section*{APPENDIX B}  
\section*{Derivation of the Approximate MAP Detector Based on the Smoother--Detector Structure}
\label{sec:app_a}

The approximate MAP detector based on the smoother-detector structure in \refeq{eq:estdetappr} is rewritten as
\begin{IEEEeqnarray}{rCl}\label{eq:estdet_appr_a}
\hat{\bm c}_{k} &\propto&  {\arg\underset{ \bck{k} } \max} \;\exp\!\left\{-\frac{|\ck{k}{1}|^{2}+|\ck{k}{2}|^{2}}{N_{0}}\right\} \underset{\bThetak{k}}\int \exp\!\left\{ \Re\left[ \frac{2}{N_{0}}\rk{k}{1}{\ck{k}{1}}^{*} \right.\right. \nonumber\\ &&\left.\left. \cdot e^{-\jmath \thetak{k}{1,1}}  +\frac{2}{N_{0}}\rk{k}{2}{\ck{k}{2}}^{*}e^{-\jmath\thetak{k}{2,1}}  -\frac{2}{N_{0}}\ck{k}{2}{\ck{k}{1}}^{*} \right.\right. \nonumber\\ &&\left.\left. \cdot e^{-\jmath(\thetak{k}{1,1} - \thetak{ k}{2,1})}\right]\right\} \N(\bThetak{k};\bThetahatk{k},\mathbf{\hat{P}}_{k}) \tr d \bThetak{k}  \\
\label{eq:estdet_appr_b}
&\approx&   {\arg\underset{ \bck{k} } \max}\;\exp\!\left\{-\frac{|\ck{k}{1}|^{2}+|\ck{k}{2}|^{2}}{N_{0}}\right\} \underset{\bThetak{k}}\int \exp\!\left\{ \Re\left[ \frac{2}{N_{0}}\rk{k}{1}{\ck{k}{1}}^{*} \right.\right. \nonumber\\ &&\left.\left. \cdot e^{-\jmath \thetak{k}{1,1}}  +\frac{2}{N_{0}}\rk{k}{2}{\ck{k}{2}}^{*}e^{-\jmath\thetak{k}{2,1}}  -\frac{2}{N_{0}}\ck{k}{2}{\ck{k}{1}}^{*} \right.\right. \nonumber\\ &&\left.\left. \cdot e^{-\jmath(\thetak{k}{1,1} - \thetak{ k}{2,1})}\right]\right\} \exp\!\left\{ \Re\left[ \frac{  e^{\jmath \hat{\theta}_{k}^{(1,1)}}}{P_{k,1}^{(1,1)}} e^{-\jmath \thetak{k}{1,1}}  \right.\right. \nonumber\\ &&\left.\left.  +\frac{  e^{\jmath \hat{\theta}_{k}^{(2,1)}}}{P_{k,1}^{(2,1)}}e^{-\jmath\thetak{k}{2,1}}  - \left| \uttk{k}{1,2}\right| e^{-\jmath(\thetak{k}{1,1} - \thetak{ k}{2,1})}\right]\right\} \tr d \bThetak{k} \\
&\approx&   {\arg\underset{ \bck{k} } \max} \; B_{\tr{temp}}^{(1)},
\end{IEEEeqnarray}
where, in \refeq{eq:estdet_appr_b}, we approximate the bivariate Gaussian pdf as a bivariate Tikhonov distribution \cite{mardia07}. We simplify $B_{\tr{temp}}^{(1)} $ as
\begin{IEEEeqnarray}{rCl}\label{eq:estdet_appr_d}
&&B_{\tr{temp}}^{(1)} \nonumber\\
&&\triangleq \exp\!\left\{-\frac{|\ck{k}{1}|^{2}+|\ck{k}{2}|^{2}}{N_{0}}\right\} \exp\!\left\{ \Re\left[ \uk{k}{2,1}e^{-\jmath\thetak{k}{2,1}} \right]\right\} \nonumber\\ && \cdot
\underset{\bThetak{k}}\int \exp\!\left\{ \Re\left[ \uk{k}{1,1}e^{-\jmath \thetak{k}{1,1}}  - \utk{k}{1,2}e^{-\jmath(\thetak{k}{1,1} - \thetak{ k}{2,1})}\right]\right\} \tr d \bThetak{k}\IEEEeqnarraynumspace\\
\label{eq:estdet_appr_e}
&&\propto  \exp\!\left\{-\frac{|\ck{k}{1}|^{2}+|\ck{k}{2}|^{2}}{N_{0}}\right\} \mbox{I}_{0}\left(|\uk{k}{1,1} - \utk{k}{1,2}e^{-\jmath\thetak{k}{2,1}} |\right) \nonumber\\&&  \cdot \underset{\bThetak{k}}\int \exp\!\left\{ \Re\left[ \uk{k}{2,1}e^{-\jmath\thetak{k}{2,1}} \right]\right\}  \tr d \bThetak{k} \\
\label{eq:estdet_appr_f}
&&\approx  \exp\!\left\{-\frac{|\ck{k}{1}|^{2}+|\ck{k}{2}|^{2}}{N_{0}}\right\}\mbox{I}_{0}\left(||\uk{k}{1,1}| -  |\utk{k}{1,2}| |\right)  \mbox{I}_{0}\left(|\uk{k}{2,1}|\right) \nonumber\\ &&  \\
\label{eq:estdet_appr_g}
&&\approx  \exp\!\left\{-\frac{|\ck{k}{1}|^{2}+|\ck{k}{2}|^{2}}{N_{0}}\right\} \mbox{I}_{0}\left(|\uk{k}{1,1}| + |\uk{k}{2,1}| - |\utk{k}{1,2}| \right),\nonumber\\
\end{IEEEeqnarray}
where we use the definitions from \refeq{eq:udefs}  in \refeq{eq:estdet_appr_d} and the Tikhonov pdf normalization constant in \refeq{eq:estdet_appr_e}. In \refeq{eq:estdet_appr_f}, we use the approximation from \refeq{eq:ya2_apprx} where it is assumed that the difference between the predicted and updated estimates and the states is very small. Finally, in \refeq{eq:estdet_appr_g} we approximate $\mbox{I}_{0}(x) \approx e^{x}$ for large $x$ and assume $|\uk{k}{1,1}| -  |\utk{k}{1,2}| > 0$. The generalization presented in \refeq{eq:estdetappr_gen} is a straightforward extension of the computations from \refeq{eq:estdet_appr_a}-\refeq{eq:estdet_appr_g} for arbitrary values of $N_{\tr t}$ and $N_{\tr r}$.

\section*{APPENDIX C}  
\section*{Derivation of the Factorized pdfs for the VB Framework}
\label{sec:app_c}

From \refeq{eq:pthetapc_update}, the factorized pdf of $\bbTheta$ is derived as
\begin{IEEEeqnarray}{rCl}
&& q_{\boldsymbol{\theta}}(\bbTheta) =  C_{\boldsymbol{\theta}} p(\bbTheta)\prod_{k=1}^{L} \exp\!\left\{ {C_{\tr{temp}}^{(2)}}\right\}, \mbox{where} \nonumber\\
&& C_{\tr{temp}}^{(2)}  \triangleq  \mathbb{E}_{q_{\mathbf{c}}} \left\{\log p(\brk{k}|\bck{k},\bThetak{k})\right\}\nonumber\\
&& =  \mathbb{E}_{q_{\mathbf{c}}} \left\{\log \overset{N_{\tr r}}{\underset{n=1}{\prod}} p(\rk{k}{n}|\bck{k},\bThetak{k})\right\} ,\nonumber\\
&& =   -\frac{1}{N_{0}}\sum_{n=1}^{N_{\tr r}} \,\mathbb{E}_{q_{\mathbf{c}}}\left(\rk{k}{n}{\rk{k}{n}}^{*} \right.\nonumber\\&&\left. -\overset{N_{\tr t}}{\underset{m=1}{\sum}}\left(\left|\ck{k}{m}\right|^{2} +\underset{\underset{l\neq m}{l=1}}{\overset{N_{\tr t}}{\sum}}\ck{k}{m}{\ck{k}{l}}^{*}e^{\jmath\left(\thetak{k}{m,n}-\thetak{k}{l,n}\right)}\right)\right.\nonumber\\
&&\left.-{\rk{k}{n}}^{*}\overset{N_{\tr t}}{\underset{m=1}{\sum}}\ck{k}{m}e^{\jmath\thetak{k}{m,n}}-\rk{k}{n}\overset{N_{\tr t}}{\underset{m=1}{\sum}}{\ck{k}{m}}^{*}e^{-\jmath \thetak{k}{m,n}}\right) \nonumber\\
\label{eq:q_theta_derivation_a}
&=& -\frac{1}{N_{0}}\overset{N_{\tr r}}{\underset{n=1}{\sum}}\left(\rk{k}{n}{\rk{k}{n}}^{*}+\overset{N_{\tr t}}{\underset{m=1}{\sum}}\left(\left|\underline{c}_{k}^{(m)}\right|^{2} \right.\right. \nonumber\\ && \left.\left.  +\underset{\underset{l\neq m}{l=1}}{\overset{N_{\tr t}}{\sum}}\underline{c}_{k}^{(m)}{\underline{c}_{k}^{(m)}}^{*}e^{\jmath\left(\thetak{k}{m,n}-\thetak{k}{m,n}\right)}\right)-{\rk{k}{n}}^{*}\overset{N_{\tr t}}{\underset{m=1}{\sum}}\underline{c}_{l}^{(m)}\right.\nonumber\\
&& \left. \cdot e^{\jmath\thetak{k}{m,n}}-\rk{k}{n}\overset{N_{\tr t}}{\underset{m=1}{\sum}}{\underline{c}_{l}^{(m)}}^{*}e^{-\jmath\thetak{k}{m,n}}-\overset{N_{\tr t}}{\underset{m=1}{\sum}}{\mathrm{Var}}_{q_{\textbf{c}}}\ck{k}{m} \right). \\
&=& C_{\boldsymbol{\theta}}p(\bbTheta)\prod_{k=1}^{L} \exp\!\left\{ \overset{N_{\tr r}}{\underset{n=1}{\sum}}\frac{-\left|\rk{k}{n}-\overset{N_{\tr t}}{\underset{m=1}{\sum}}\underline{c}_{k}{(m,n)}e^{\jmath\thetak{k}{m,n}}\right|^{2}}{N_{0}}\right\} \nonumber\\
\label{eq:q_theta_derivation_b}
&=&  C_{\boldsymbol{\theta}}P(\bbTheta)\prod_{k=1}^{L}p(\brk{k}|\bThetak{k},\mathbf{\underline{c}}_{k}),\\
 \label{eq:q_theta_derivation_c}
&=&  C_{\boldsymbol{\theta}}p(\bbr,\bbTheta|\mathbf{\underline{c}}).
\end{IEEEeqnarray}
In \refeq{eq:q_theta_derivation_a}, $\underline{c}_{l}^{(m)}$ is defined as in \refeq{eq:q_theta}, and for obtaining the result in \refeq{eq:q_theta_derivation_b} we apply the approximation ${\mathrm{Var}}_{q_{\textbf{c}}}  =  \mathbb{E}_{q_{\mathbf{c}}} \left\{ \ck{k}{m}{\ck{k}{m}}^{*}-\underline{c}_{k}^{(m)}{\underline{c}_{k}^{(m)}}^{*} \right\} \approx  0$. The constant $C_{\boldsymbol{\theta}}$ is the pdf normalizing factor such that $q_{\boldsymbol{\theta}}$ integrates to unity. Its value can be determined as
 \begin{IEEEeqnarray}{rCl}\label{eq:c_theta}
C_{\boldsymbol{\theta}}  &=&  \frac{1}{p(\bbr|\mathbf{\underline{c}})}. \nonumber
\end{IEEEeqnarray}
Plugging $C_{\boldsymbol{\theta}}$ in \refeq{eq:q_theta_derivation_c}, the factorized pdf of $\bbTheta$ is obtained as
\begin{IEEEeqnarray}{rCl}\label{eq:q_thetafinal}
q_{\boldsymbol{\theta}}  =  p(\bbTheta|\bbr,\underline{\mathbf{c}}).
\end{IEEEeqnarray}

From \refeq{eq:pthetapc_update}, the factorized pmf of $\bbc$ is derived as
\begin{IEEEeqnarray}{rCl}\label{eq:q_symbol_temp}
&& q_{\mathbf{c}} =  C_{\boldsymbol{c}}\prod_{k=1}^{L} P(\bck{k}) \exp\!\left\{{C_{\tr{temp}}^{(3)}}\right\} , \mbox{ where}\\
 && C_{\tr{temp}}^{(3)} \triangleq {\underset{\boldsymbol{\theta}}{\int}q_{\boldsymbol{\theta}}}\log P(\brk{k}|\bck{k},\bThetak{k})\tr d \bThetak{k} \nonumber\\
 && =   \mathbb{E}_{q_{\boldsymbol{\theta}}} \log \overset{N_{\tr r}}{\underset{n=1}{\prod}}p(\rk{k}{n}|\bck{k},\bThetak{k})  \nonumber\\
 && =  -\frac{1}{N_{0}}\sum_{n=1}^{N_{\tr r}} \mathbb{E}_{q_{\boldsymbol{\theta}}} \left|\rk{k}{n}-\overset{N_{\tr t}}{\underset{m=1}{\sum}}\ck{k}{n}e^{\jmath\thetak{k}{k}}\right|^{2}\nonumber\\
 && =  -\frac{1}{N_{0}}\overset{N_{\tr r}}{\underset{n=1}{\sum}}\left\{ \rk{k}{n}{\rk{k}{n}}^{*}
 \left(\left|\ck{k}{m}\right|^{2}\right.\right. \nonumber\\ && \left.\left.  +\overset{N_{\tr t}}{\underset{m=1}{\sum}} \left(1+\mathcal{\mathbb{E}}_{q_{\boldsymbol{\theta}}}(\thetak{k}{m,n}-\thetahatk{k}{m,n})^{2}\right) +\underset{\underset{l\neq m}{l=1}}{\overset{N_{\tr t}}{\sum}}\ck{k}{m}{\ck{k}{l}}^{*}   \right.\right.\nonumber\\
 && \left.\left. \cdot e^{\jmath(\thetahatk{k}{m,n}-\thetahatk{k}{l,n})}(1+\mathcal{\mathbb{E}}_{q_{\boldsymbol{\theta}}}(\thetak{k}{m,n}-\thetahatk{k}{m,n})(\thetak{k}{l,n}-\thetahatk{k}{l,n}))\right. \right.\nonumber\\
 && \left.\left. -{\rk{k}{n}}^{*}\overset{N_{\tr t}}{\underset{m=1}{\sum}}\ck{k}{m}(n)e^{\jmath\thetahatk{k}{m,n}}-\rk{k}{n}\overset{N_{\tr t}}{\underset{m=1}{\sum}}{\ck{k}{m}}^{*}e^{-\jmath\thetahatk{k}{m,n}}\right\}\right.,\nonumber\\
 && =  -\frac{1}{N_{0}} \left\{\overset{N_{\tr r}}{\underset{n=1}{\sum}} {\left|\rk{k}{n}-\overset{N_{\tr t}}{\underset{m=1}{\sum}}\ck{k}{m}e^{\jmath\thetahatk{k}{m,n}}\right|^{2}} -\overset{N_{\tr t}}{\underset{m=1}{\sum}}{\left|\ck{k}{m}\right|^{2}P_{k,n}^{(m,m)} } \right.\nonumber\\
 \label{eq:q_symbol_temp1}
 && \left.-\overset{N_{\tr t}}{\underset{m=1}{\sum}}\underset{\underset{l\neq m}{l=1}}{\overset{N_{\tr t}}{\sum}}{\ck{k}{m}{\ck{k}{l}}^{*}P_{k,n}^{(m,l)}e^{\jmath(\thetahatk{k}{m,n} - \jmath\thetahatk{k}{l,n})}} \right\}.
\end{IEEEeqnarray}
Upon plugging the result from \refeq{eq:q_symbol_temp1} into  \refeq{eq:q_symbol_temp} we obtain the result in \refeq{eq:q_symbol}.

\bibliographystyle{IEEEbib}

\end{document}